\documentclass[superscriptaddress,floatfix,aps,prd,showpacs,twocolumn]{revtex4-1}

\usepackage{graphicx}
\usepackage{color}
\usepackage{amsfonts,amsmath,amssymb,amsthm}
\usepackage[utf8]{inputenc}
\usepackage[T1]{fontenc}
\usepackage{lmodern}
\usepackage[bookmarks, bookmarksopen, bookmarksnumbered]{hyperref}
\usepackage[right]{rotating}

\newcommand{\diff}{\mathrm{d}}
\newcommand{\codename}[1]{\texttt{#1}}

\newcommand{\GRHydro}{\codename{GRHydro}~}

\graphicspath{{./}}

\begin{document}
\title{Neutron Star instabilities in full General Relativity using a $\Gamma=2.75$ ideal fluid}
\date{\today}

\author{Roberto \surname{De Pietri}}
\affiliation{Parma University and INFN Parma, via G.P. Usberti 7/A, I-43124 Parma (PR), Italy}
\author{Alessandra \surname{Feo}}
\affiliation{Parma University and INFN Parma, via G.P. Usberti 7/A, I-43124 Parma (PR), Italy}
\author{Luca \surname{Franci}}
\affiliation{Parma University and INFN Parma, via G.P. Usberti 7/A, I-43124 Parma (PR), Italy}
\author{Frank \surname{L\"offler}}
\affiliation{Center for Computation \& Technology, Louisiana State University, Baton Rouge, LA 70803 USA}

\begin{abstract}
We present results about the effect of the use of a stiffer equation of
state, namely the ideal-fluid $\Gamma=2.75$ ones, on the dynamical
bar-mode instability in rapidly rotating polytropic models of neutron
stars in full General Relativity.  We determine the change on the
critical value of the instability parameter $\beta$ for the emergence
of the instability when the adiabatic index $\Gamma$ is changed from 2
to 2.75 in order to mimic the behavior of a realistic equation of
state.  In particular, we show that the threshold for the onset of the
bar-mode instability is reduced by this change in the stiffness and
give a precise quantification of the change in value of the critical
parameter $\beta_c$. We also extend the analysis to lower values of
$\beta$ and show that low-beta shear instabilities are present also in
the case of matter described by a simple polytropic equation of state.
\end{abstract}


\maketitle

\section{Introduction}
\label{sec:intro}

Non-axisymmetric deformations of rapidly rotating self-gravitating
bodies are a rather generic phenomenon in nature and could appear in a
variety of astrophysical scenarios like stellar core
collapses~\cite{Shibata:2004kb,Ott:2005gj}, accretion-induced collapse
of white dwarfs~\cite{Burrows:2007yx}, or the merger of two neutron
stars~\cite{Shibata:2003ga,Shibata:2005ss}. Over the years, a
considerable amount of work has been devoted to the search of unstable
deformations that, starting from an axisymmetric configuration, can
lead to the formation of highly deformed rapidly rotating massive
objects~\cite{Shibata:2000jt,Baiotti:2006wn,Kruger:2009nw,Kastaun:2010vw,Lai:1994ke}.
Such deformations would lead to an intense emission of high-frequency
gravitational waves (i.e. in the kHz range), potentially detectable on
Earth by next-generation gravitational-wave detectors such as Advanced
LIGO~\cite{Harry:2010}, Advanced VIRGO and KAGRA~\cite{KAGRA:2012} in
the next decade~\cite{LIGOVIRGO:2013}.

From the observational point of view, it is import to get any insight on 
the possible astrophysical scenarios where such instabilities (unstable deformation)
are present.
It is well known that rotating neutron stars are subject to
non-axisymmetric instabilities for non-radial axial modes with
azimuthal dependence $\mathrm{e}^{i m \phi}$ (with $m = 1,2,\ldots$)
when the instability parameter $\beta \equiv T/|W|$ (i.e. the ratio
between the kinetic rotational energy $T$ and the gravitational
potential energy $W$) exceeds a critical value $\beta_c$. The
instability parameter plays an important role in the study of the
so-called dynamical bar-more instability, i.e. the $m=2$ instability
which takes place when $\beta$ is larger than a threshold
~\cite{Baiotti:2006wn}.  Previous results for the onset of
the classical bar-mode instability have already showed that the
critical value $\beta_c$ for the onset of the instability is not an
universal quantity and it is strongly influenced by the rotational
profile~\cite{Shibata:2003yj,KarinoEriguchi03}, by relativistic
effects~\cite{Shibata:2000jt,Baiotti:2006wn}, and, in a quantitative way, by the
compactness~\cite{Manca:2007ca}. 

However, up to now, significant evidence of their presence when 
realistic Equation of State (EOS) are consider is still missing. For example in \cite{Corvino:2010},
using the unified SLy EOS~\cite{Douchin01}, was shown the presence of shear-instability
but no sign of the classical bar-mode instability and of its critical behavior
have been found. The main aim of the present work is to get more insight
on the behavior of the classical bar-mode instability when the matter is
described by a stiffer more realistic EOS.  The
investigation in the literature on its dependence on the stiffness of EOS usually focused 
on the values of $\Gamma$
(i.e. the adiabatic index of a polytropic EOS) in the range between
$\Gamma=1$ and $\Gamma=2$~\cite{Lai:1994ke,2007PhRvD..76b4019Z,Kastaun:2010vw}, 
while the expected value for a real neutron
star is more likely to be around $\Gamma=2.75$ at least in large portions
of the interior. Such a choice for the EOS has already been
implemented in the past~\cite{Oechslin2007aa}, even quite recently~\cite{Giacomazzo:2013uua}, 
with the aim of maintaining the simplicity
of a polytropic EOS and yet obtaining properties that resemble a more
realistic case. Indeed, as it is shown in Fig.~\ref{fig:EOSs}, a
polytropic EOS with $K=30000$ and $\Gamma=2.75$ is qualitatively
similar to the Shen proposal~\cite{shen98,shen98b} in the density
interval between $2 \times 10^{13} \text{g/cm}^3$ and $10^{15}
\text{g/cm}^3$. For the sake of completeness, in Fig.~\ref{fig:EOSs}
we also report the behavior of the $\Gamma=2$ polytrope used
in~\cite{Baiotti:2006wn,Manca:2007ca} and of the unified SLy
EOS~\cite{Douchin01} which describes the high-density cold (zero
temperature) matter via a Skyrme effective potential for the
nucleon-nucleon interactions~\cite{Corvino:2010}.

The organization of this paper is as follows. In Sect.~\ref{sec:setup}
we describe the main properties of the relativistic stellar models we
investigated and briefly review the numerical setup used for their
evolutions.  In Sect.~\ref{sec:results} we present and discuss our
results, showing the features of the evolution for models that lie
both above and below the threshold for the onset of the bar-mode instability and
quantifying the effects of the compactness on the onset of the
instability.  Conclusions are finally drawn in
Sect.~\ref{sec:conclusions}.  Throughout this paper we use a
space-like signature $-,+,+,+$, with Greek indices running from 0 to
3, Latin indices from 1 to 3 and the standard convention for summation
over repeated indices. Unless otherwise stated, all quantities are
expressed in units in which $c=G=M_\odot=1$.

\section{Initial models and Numerical setup}
\label{sec:setup}

In this work we solve the Einstein's field equations
\begin{equation}
 G_{\mu\nu} = 8\pi T_{\mu\nu} \, , 
\end{equation}
where $G_{\mu\nu}$ is the Einstein tensor of the four-dimensional
metric $g_{\mu\nu}$ and $T^{\mu\nu}$ is the stress-energy 
tensor of an ideal fluid. This can be parametrized as 
\begin{equation}
 T^{\mu\nu} = \rho\left(1+\epsilon+\frac{P}{\rho}\right)u^\mu u^\nu + Pg^{\mu\nu} \, ,
\end{equation}
where $\rho$ is the rest-mass density, $\epsilon$ is the specific internal energy of the matter,
$P$ is the pressure and $u^\mu$ is the matter $4$-velocity.
The evolution equations for the matter follow from the conservation laws for
the energy-momentum tensor $\nabla_\mu T^{\mu\nu}=0$ and the baryon number 
$\nabla_\mu(\rho u^\mu)=0$, closed by an EOS of the type $P=P(\rho,\epsilon)$.

\begin{figure}
\vspace{-4mm} 
\begin{centering}
  \includegraphics[width=0.45\textwidth]{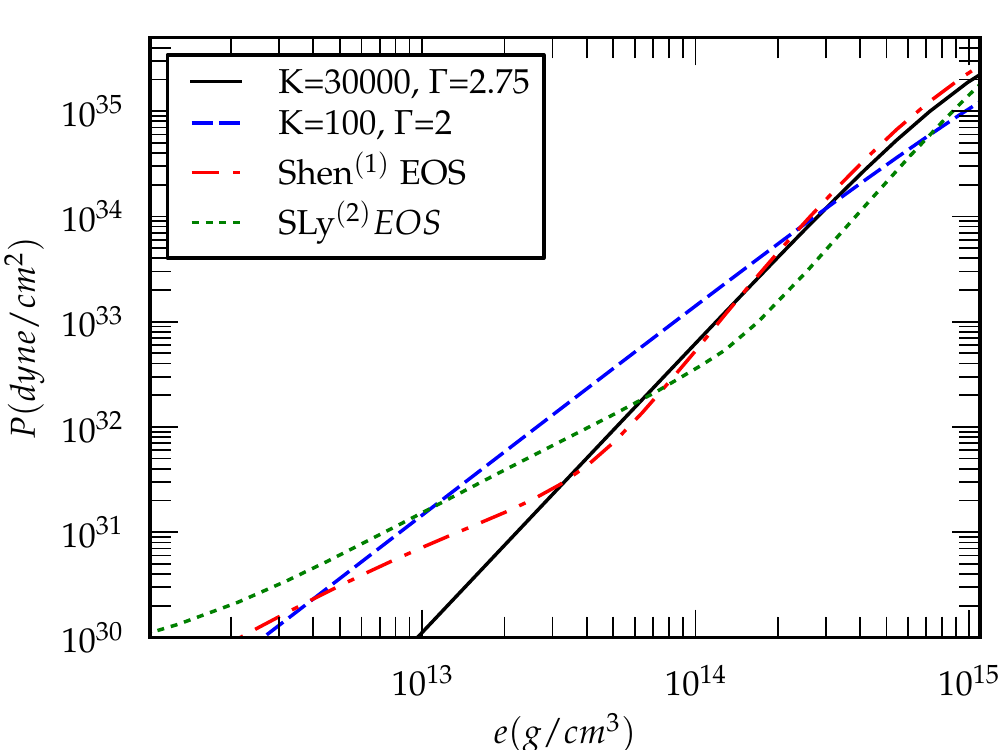}\\
\end{centering}
\vspace{-3mm}
\caption{Diagram of the pressure $P$ vs. the energy density $e$ for two polytropic
         EOSs and two {\it realistic} EOS for nuclear matter,
         namely: (1) the Shen proposal~\cite{shen98,shen98b};
         (2) the unified SLy prescription~\cite{Douchin:2001sv}.
\label{fig:EOSs}}
\end{figure}

In order to generate the initial data we evolve in this work, we use a
$\Gamma$-type EOS of the form
\begin{equation}
 P = K\rho^\Gamma \, ,
 \label{eq:eosP_Gamma}
\end{equation} 
where the following relation between $\epsilon$ and $\rho$ holds:
$\epsilon=K \rho^{(\Gamma-1)} / (\Gamma-1)$. On the other hand, the
evolution is performed using the so-called {\it ideal-fluid}
($\Gamma$-law) EOS
\begin{equation}
 P = (\Gamma-1) \rho \epsilon \, ,
 \label{eq:eosP_Ideal}
\end{equation} 
that allows for increase of the internal energy, by shock heating, if shocks are presents.
We have chosen the EOS polytropic parameters to be $\Gamma=2.75$ for the adiabatic index
and $K=30000$ for the polytropic constant. This choice of parameters has
the property to closely reproduce the behavior of the Shen EOS in the interior of a
real neutron star (see Fig.~\ref{fig:EOSs}). We note that the
choice we make here is different from the one of our previous
studies~\cite{DePietri06,Baiotti:2006wn,Manca:2007ca}, where we used
$\Gamma=2$ and $K=100$, with the explicit intention of determining the
difference that such a change implies on the onset of the bar-mode instability.

We solve the above-mentioned set of equations using the usual $3+1$
space-time decomposition, where the space-time is foliated as a tensor
product of a three-space and a time coordinate $t$ (which is selected
to be the $x^0$ coordinate). In this coordinate system the metric can
be split as $g^{\mu\nu}=-n^\mu n^\nu + h^{\mu\nu}$, where $h^{\mu\nu}$
has only the spatial components different from zero and can be used to
define a Riemannian metric $\gamma^{ij}=h^{ij}$ on each foliation. The
vector $n^\mu$, that determines the direction normal to the
3-hypersurfaces of the foliation, is decomposed in terms of the lapse
function $\alpha$ and the shift vector $\beta^i$, such that
$n^\mu=\alpha^{-1}\cdot(1,\beta^i)$.  We also define the fluid
three-velocity $v^i$ as the velocity measured by a local zero-angular
momentum observer ($u^i=\alpha v^i - \beta^i$), while the Lorentz
factor is $\alpha u^0=\sqrt{1-\gamma_{ij}v^iv^j}$. Within this
formalism, the conservation of the baryon number suggests the use of
the conserved variable $D=\sqrt{\gamma}\alpha u^0 \rho$ with the
property that $\int D \, d^3x = \text{const}$ along the
time-evolution $t$.

\begin{figure}
\vspace{-1mm}
\begin{centering}
  \includegraphics[width=0.45\textwidth]{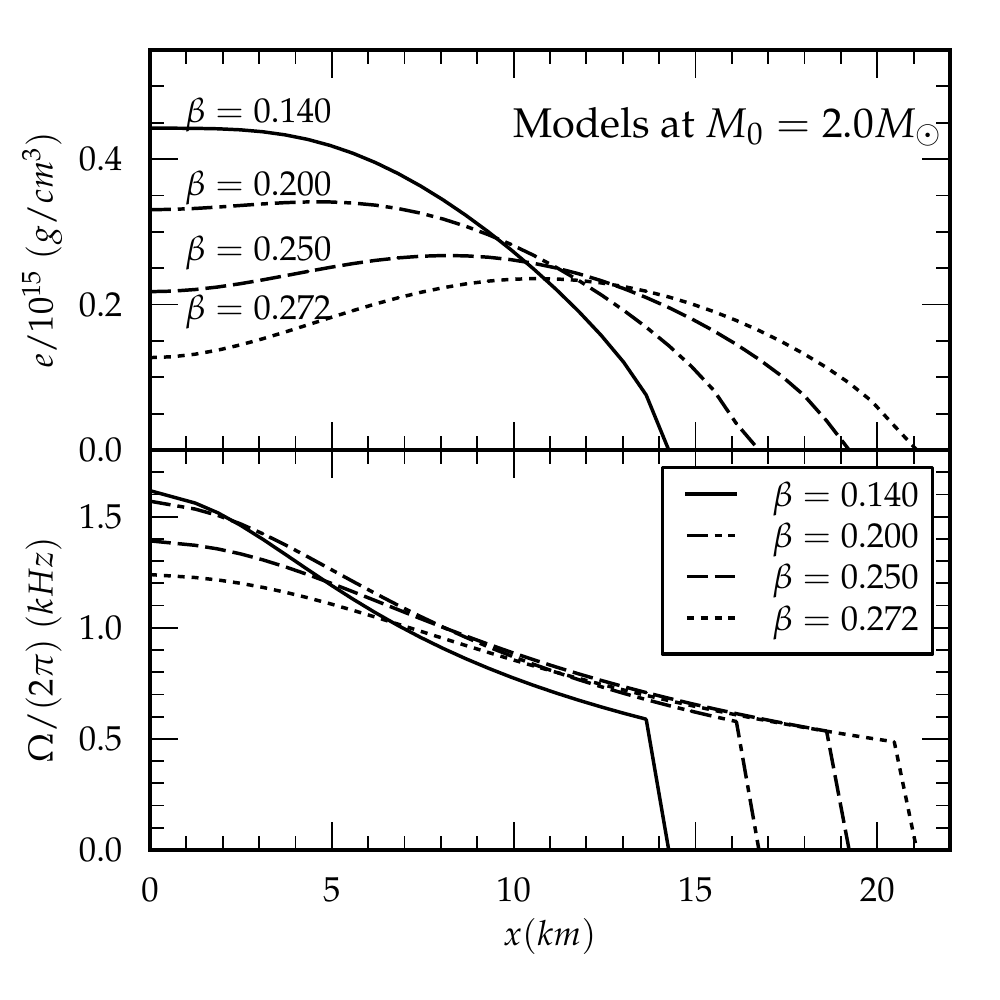}
\end{centering}
\vspace{-4mm}
\caption{Density profile (top panel) and differential 
rotation profile (bottom panel) of some representative models among all the 
ones that have been evolved in this paper.
\label{fig:densityM2}}
\end{figure}

\subsection{Initial Data}
\label{sec:models}

The initial data of our simulations are computed as stationary
equilibrium solutions for axisymmetric and rapidly rotating
relativistic stars in polar coordinates~\cite{Stergioulas95}.  In
generating these equilibrium models we assumed that the metric
describing the axisymmetric and stationary relativistic star has the
form
\begin{align}
 ds^2 = -e^{\mu + \nu} dt^2 + e^{\mu - \nu} r^2 & \sin^2\theta(d\phi -
 \omega dt^2)^2 \nonumber \\ & + e^{2 \xi} (dr^2 + r^2 d \theta^2) \, ,
\end{align}
where $\mu$, $\nu$, $\omega$, and $\xi$ are space-dependent metric
functions.  Similarly, we assume the matter to be characterized by a
non-uniform angular velocity distribution of the form
\begin{equation}
 \Omega_c - \Omega = \frac{1}{ \hat{A}^2 r_e^2 } \bigg[ \frac{ (\Omega
     - \omega) r^2 \sin^2\theta e^{-2 \nu} }{ 1 - (\Omega - \omega)^2
     r^2 \sin^2\theta e^{-2 \nu} } \bigg] \, ,
\end{equation}
where $r_e$ is the equatorial stellar coordinate radius, and the
coefficient $\hat{A}$ is the measure of the degree of the differential
rotation, which we set to be $\hat{A} = 1$, in analogy with other
works in the literature.  Once imported onto the Cartesian grid and
throughout the evolution, we compute the coordinate angular velocity
$\Omega$
on the $(x,y)$ plane as,
\begin{equation} 
\Omega = \frac{u^\phi}{u^0} = \frac{ u^y \cos\phi - u^x \sin\phi }{ u^0
  \sqrt{ x^2 + y^2 } } \, .
\end{equation}
Other characteristic quantities of the system such as the baryon mass
$M_0$, the gravitational mass $M$, the internal energy
$E_{\textrm{int}}$, the angular momentum $J$, the rotational kinetic
energy $T$, the gravitational binding energy $W$ and the instability
parameter $\beta$ are defined as~\cite{Baiotti:2006wn}:
\begin{align}
 M_0 & \equiv \int d^3x D \, , \\ M & \equiv \int d^3x (-2 T^0_0 +
 T^\mu_\mu ) \alpha \sqrt{\gamma} \, , \\ E_{\textrm{int}} & \equiv
 \int d^3x D \varepsilon \, , \\ J & \equiv \int d^3x \, T^0_{\phi}
 \alpha \sqrt{\gamma} \, , \\ T & \equiv \int d^3x \Omega T^0_{\phi}
 \alpha \sqrt{\gamma} \, , \\ W & \equiv T + E_{int} + M_0 - M \, , \\[1mm]
   \beta & \equiv T/|W| \, ,
\end{align}
where $\alpha \sqrt{\gamma}$ is the square root of the
four-dimensional metric determinant.  Notice that the definitions of
quantities such as $J$, $T$, $W$ and $\beta$ are meaningful only
in the case of stationary axisymmetric configurations and should
therefore be treated with care once the rotational symmetry is lost.
All the equilibrium models considered here have been calculated using
the relativistic polytropic EOS given in Eq.~(\ref{eq:eosP_Gamma}),
choosing $K=30000$ and $\Gamma=2.75$, in contrast
to~\cite{Baiotti:2006wn,Manca:2007ca}, where the values of $K=100$ and
$\Gamma=2$ have been used.

The initial conditions for the evolution have been generated using the
Nicholas Stergioulas' RNS code~\cite{Stergioulas95}. Any 
model can be uniquely determined (once the value of the
differential rotation parameter has been fixed to $\hat{A}=1$) by two
parameters. We decided to denote each of the generated models using
the values of the conserved baryonic mass $M_0$ and the $\beta$
parameter at $t=0$. As a consequence of this choice, in the rest 
of this paper we will refer to a particular model using the
following notation. For example, M1.5b0.270 will denote a model with a 
conserved baryonic mass $M_0 = 1.5M_\odot$ and a value of the
instability parameter $\beta=0.270$.  One of the main features of the
generated models is that, due to the high rotation, not all of them
have the maximum of the density at the center of the star. For example,
if we analyze some of the generated models with a fixed value of the
baryonic mass $M_0=2.0 \, M_\odot$ (see Fig.~\ref{fig:densityM2}), we
note that those rotating fastest have the maximum of the density at a
distance of their center which is around $15$ km.  This means that most
of the studied models are characterized by a {\it toroidal}
configuration (i.e. the maximum of the density is not on the
rotational axis). As we will see, there is no
correlation between having a toroidal configuration and being unstable
against the dynamical bar-mode instability, like in the case of polytropic 
models with $\Gamma=2$~\cite{Baiotti:2006wn}.

\begin{figure}
\vspace{-1mm}
\begin{centering}
  \includegraphics[width=0.45\textwidth]{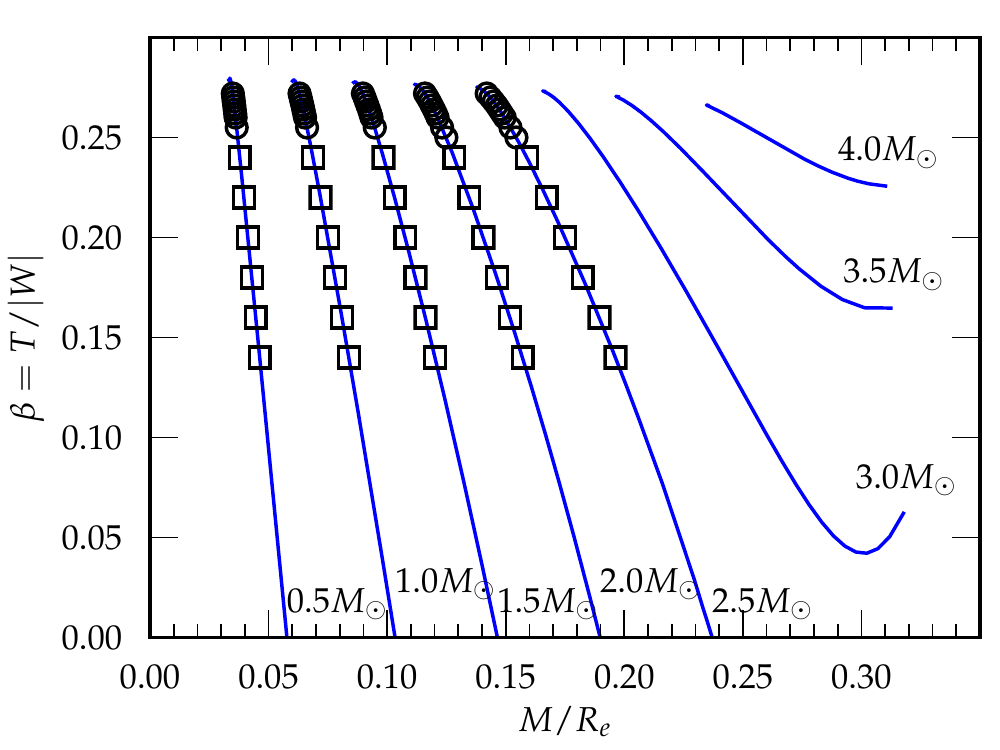}
\end{centering}
\vspace{-4mm}
\caption{Main features of the five sequences of initial models
  analyzed in the present work.  On the $x$ axis we report the
  compactness of each stellar model, while its rotation parameter
  $\beta$ is on the $y$ axis. Squares denote models that are not
  subject to the bar-mode instability while circles represent the
  ones that are bar-mode ($m$=$2$) unstable.
\label{fig:IDmodels}}
\end{figure}

\begin{figure}
\vspace{-1mm}
\begin{centering}
  \includegraphics[width=0.45\textwidth]{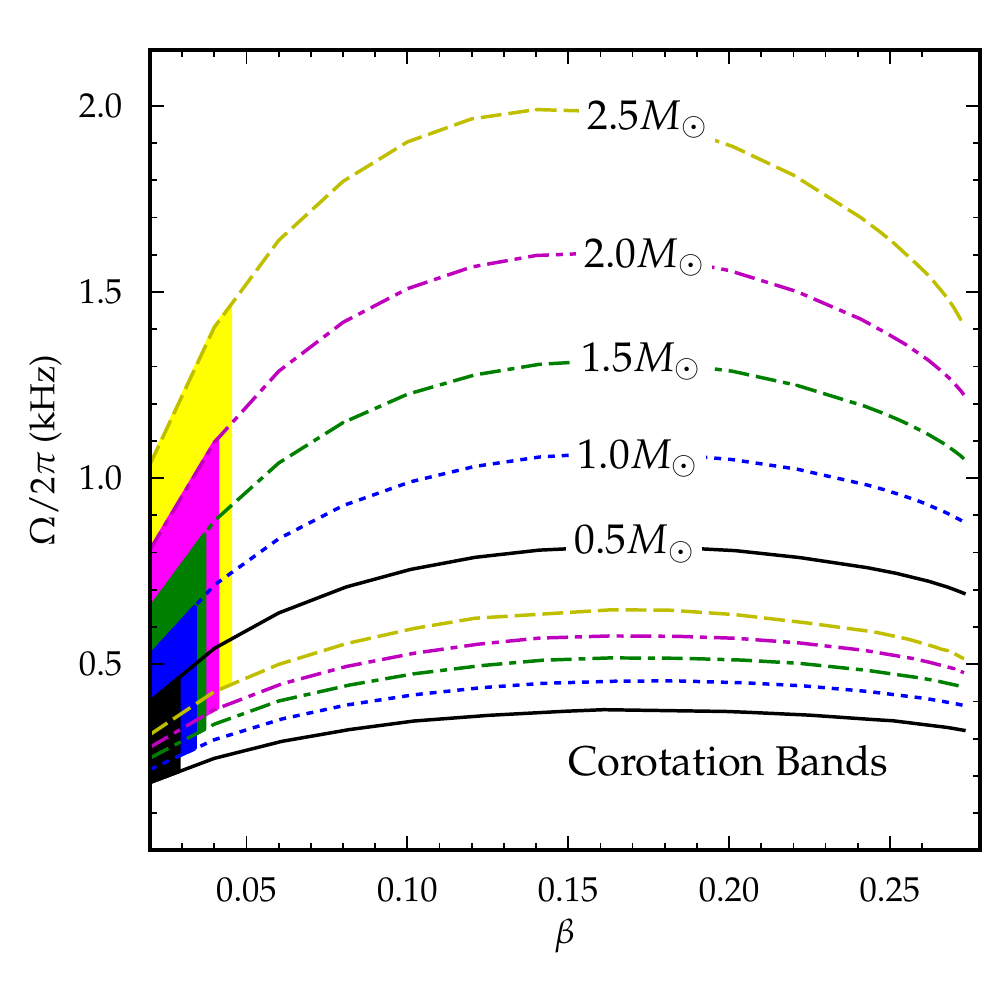}
\end{centering}
\vspace{-4mm}
\caption{Corotation bands for the five sequences of initial models studied
in the present work. For each sequence is report the 
rotation frequency on the axis (where is also reported the mass
of the sequence) and on the equator as function of the rotational 
parameter $\beta$ using continuous-black, doted-blue, 
dash-dash-doted-green, dash-doted-magenta and dashed-green line 
for the sequences of baryonic mass $0.5M\odot$,  $1.0M\odot$,
$1.5M\odot$,  $1.0M\odot$ and  $2.5M\odot$, respectively.
We also shaded, with the corresponding color, the initial part 
of the corotation bands.     
For any value of the frequency of a mode between these two lines there is 
exactly one radius inside the star which is co-rotating with the mode.
\label{fig:corotation_band}}
\end{figure}

An important issue related to the use of polytropic EOS in the construction
of the initial models is that their properties are fixed in terms of physical
scales determined by the value of the polytropic constant $K$ that can always
be set to $1$ by changing the measure units. The assertion that we are generating
a model with a giving baryonic mass $M_0$ is therefore related to the actual
value chosen for $K$. Indeed, in order to claim that the threshold for the instability
depends on the stiffness of the EOS, we need to eliminate the dependencies on the
dimensional scales and then on the chosen value of the polytropic constant $K$.
An efficient way to do
that is to extrapolate the result for $M_0 \rightarrow 0$, which
corresponds to the Newtonian limit, where the general relativistic
effects can be neglected. Indeed, using the same procedure followed
in~\cite{Manca:2007ca}, we chose five sequences of constant rest-mass
density models, namely with $M=(0.5, 1.0, 1.5, 2.0,
2.5)\,M_\odot$. Again in analogy with~\cite{Manca:2007ca}, we use a
rotational profile with $\hat A=1.0$ for all models. We restrict the
values of the instability parameter $\beta$ to the range
between $0.140$ and $0.272$ and we leave the analysis of models with
lower values to future work. The positions of all the simulated
models in terms of their compactness $M/R_e$, i.e.\ the ratio between the
gravitational mass $M$ and the equatorial radius $R_e$, and the rotational
parameter $\beta$ are reported in Fig.~\ref{fig:IDmodels}. Since the
models are differentially rotating, Fig.~\ref{fig:corotation_band}
shows the corotation bands for the five sequences of models we analyzed.

\begin{figure*}
\vspace{-4mm}
\begin{centering}
\begin{tabular}{ll}
   \hspace{-0.22cm} \includegraphics[width=0.49\textwidth]{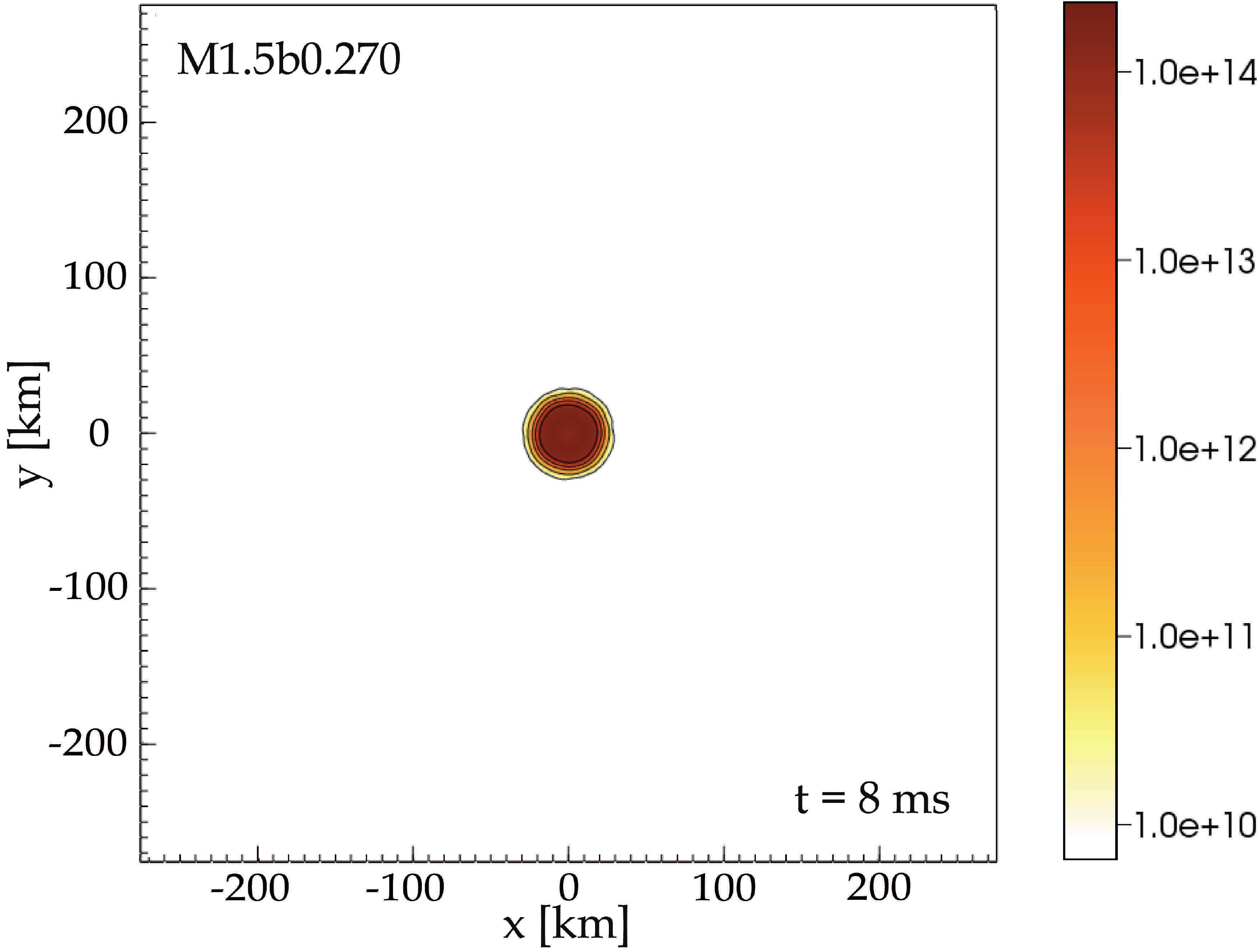} %% 8  ms
 & \hspace{-0.22cm} \includegraphics[width=0.49\textwidth]{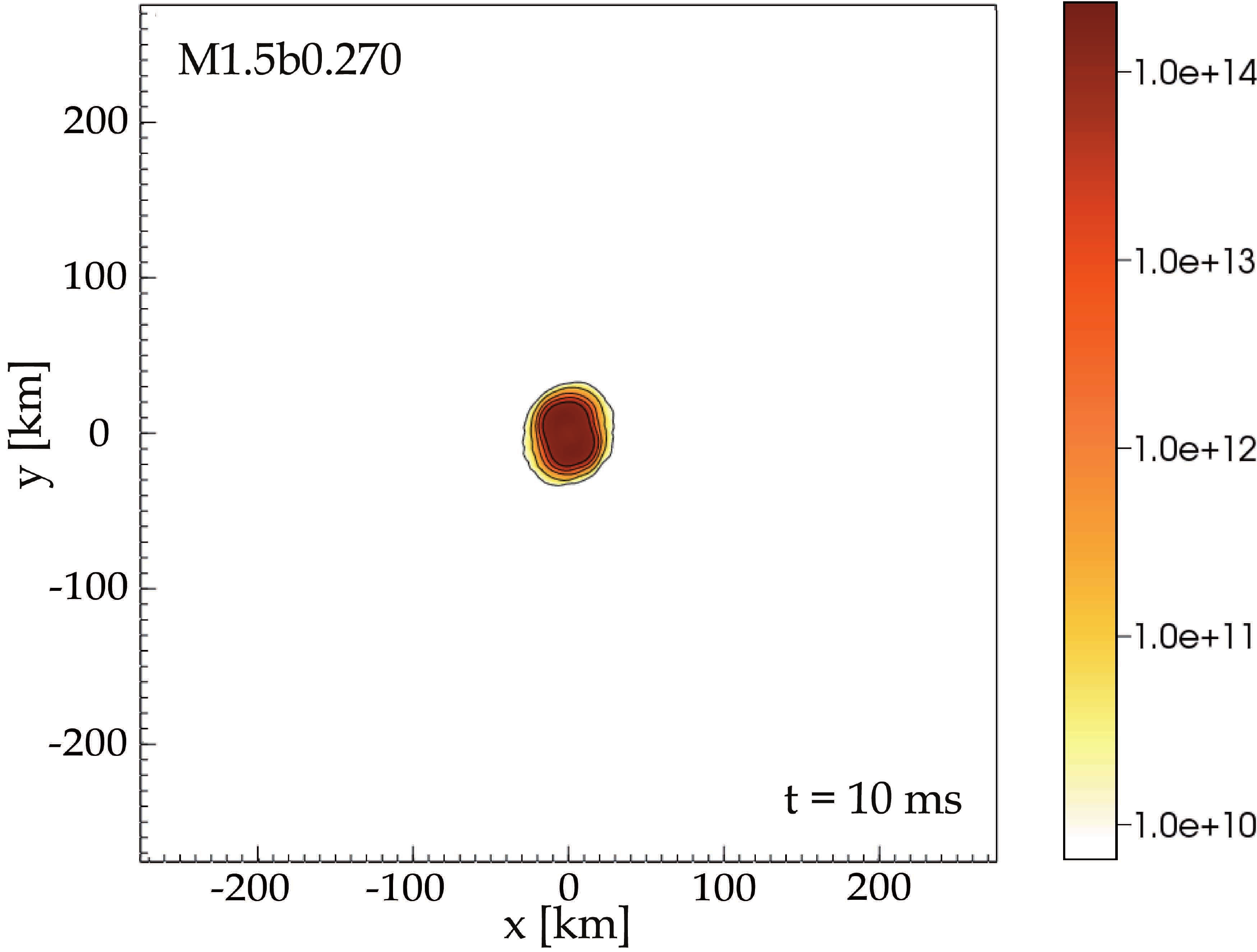} %% 10 ms
\\
   \hspace{-0.22cm} \includegraphics[width=0.49\textwidth]{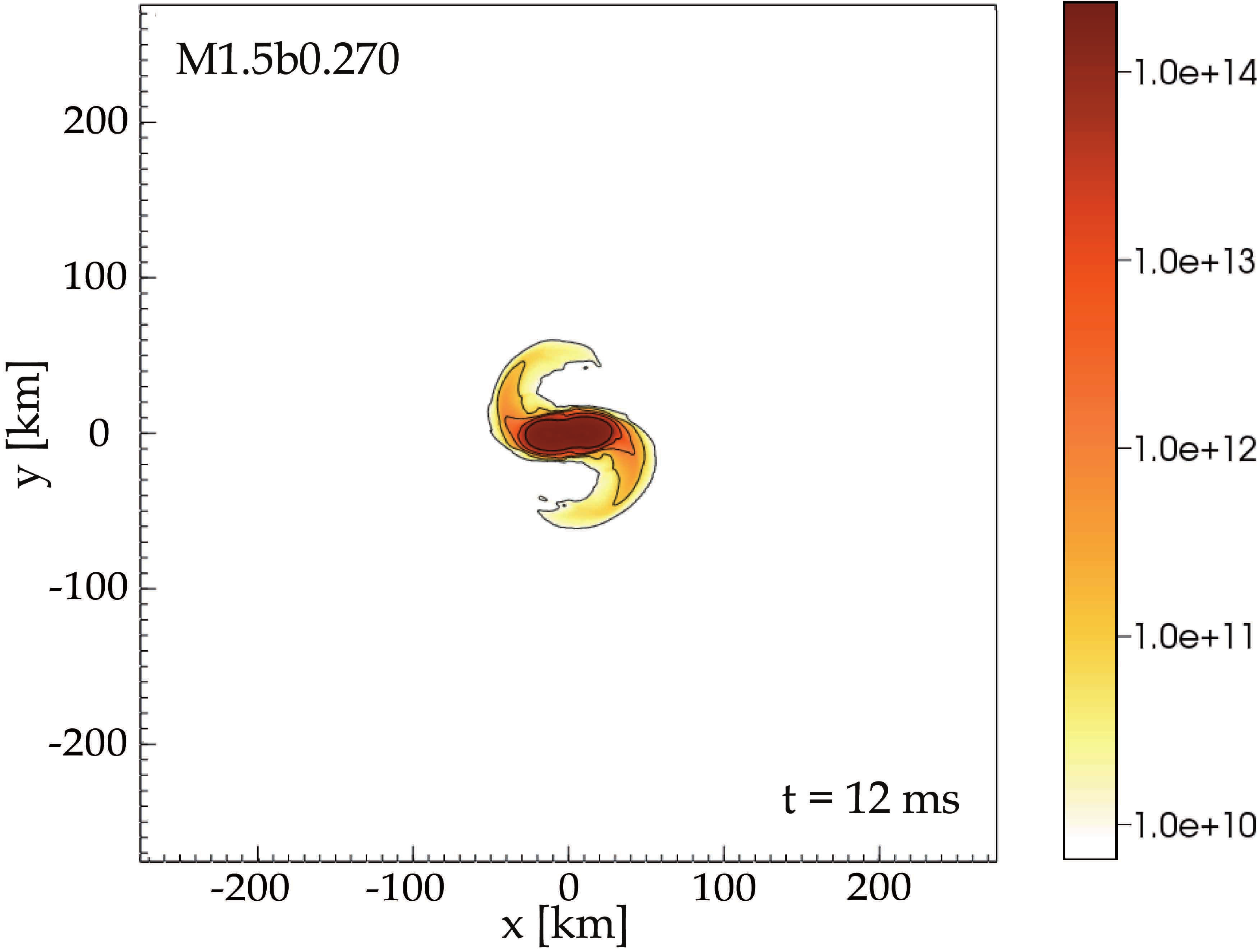} %% 12 ms
 & \hspace{-0.22cm} \includegraphics[width=0.49\textwidth]{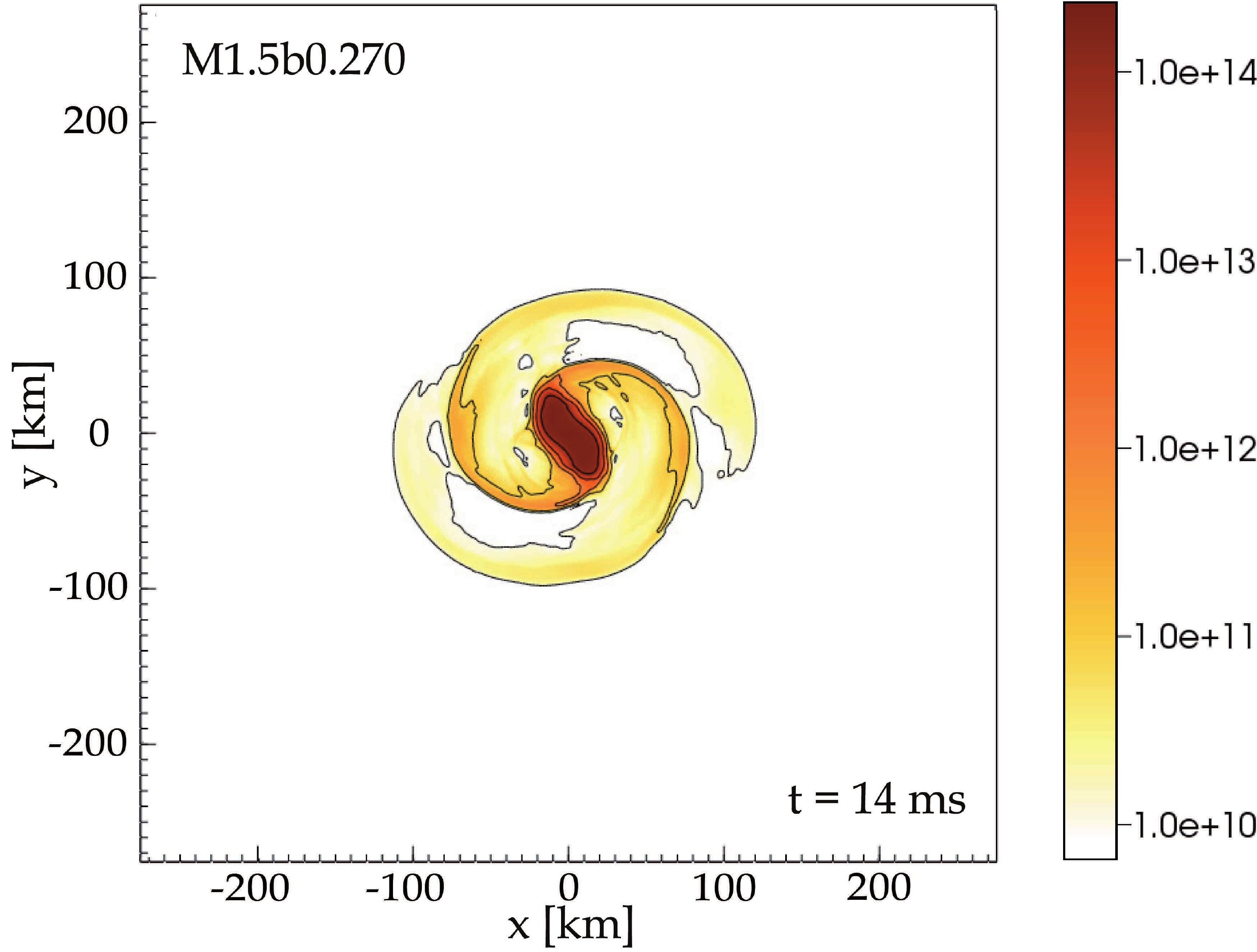} %% 14 ms
\\
   \hspace{-0.22cm} \includegraphics[width=0.49\textwidth]{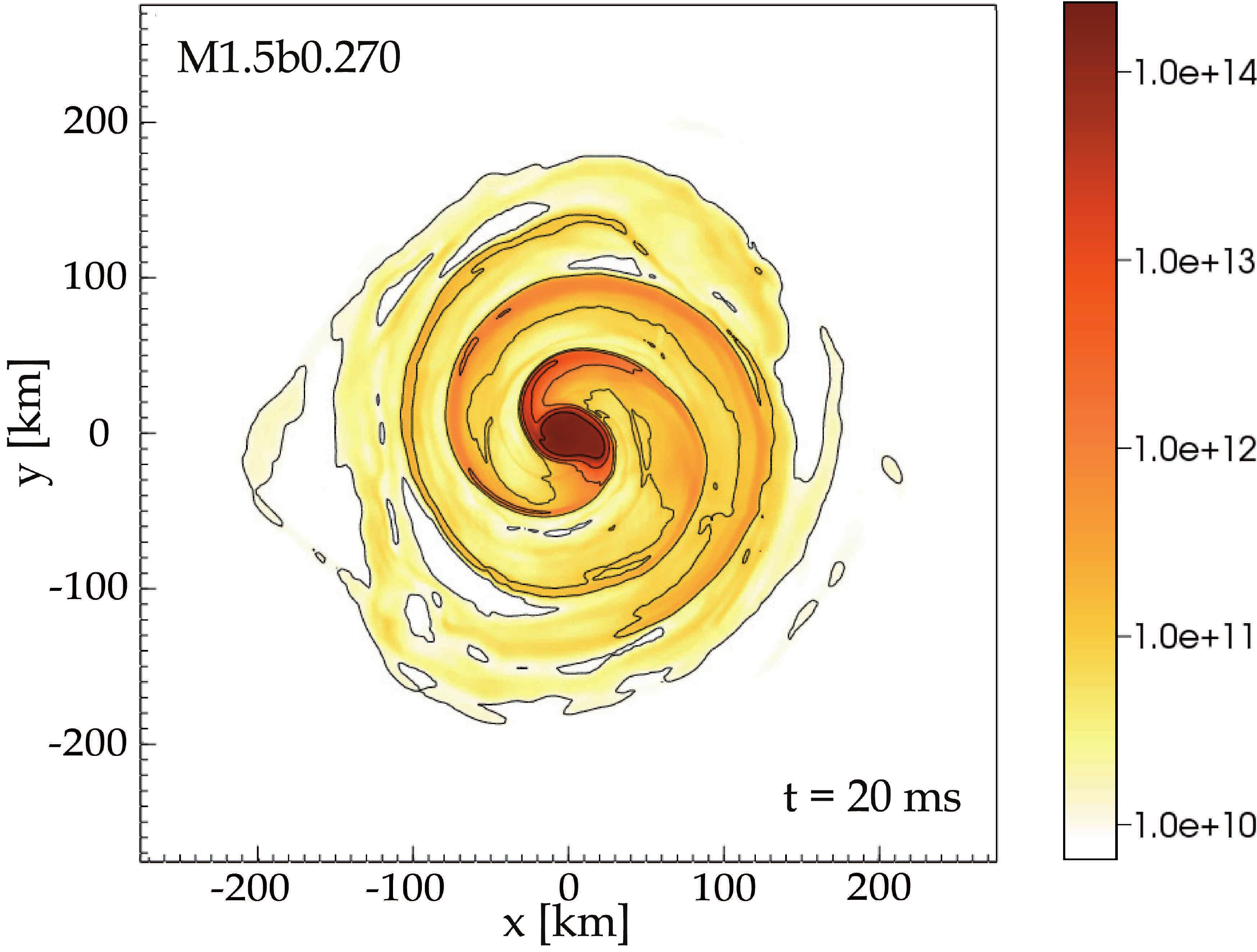} %% 20 ms
 & \hspace{-0.22cm} \includegraphics[width=0.49\textwidth]{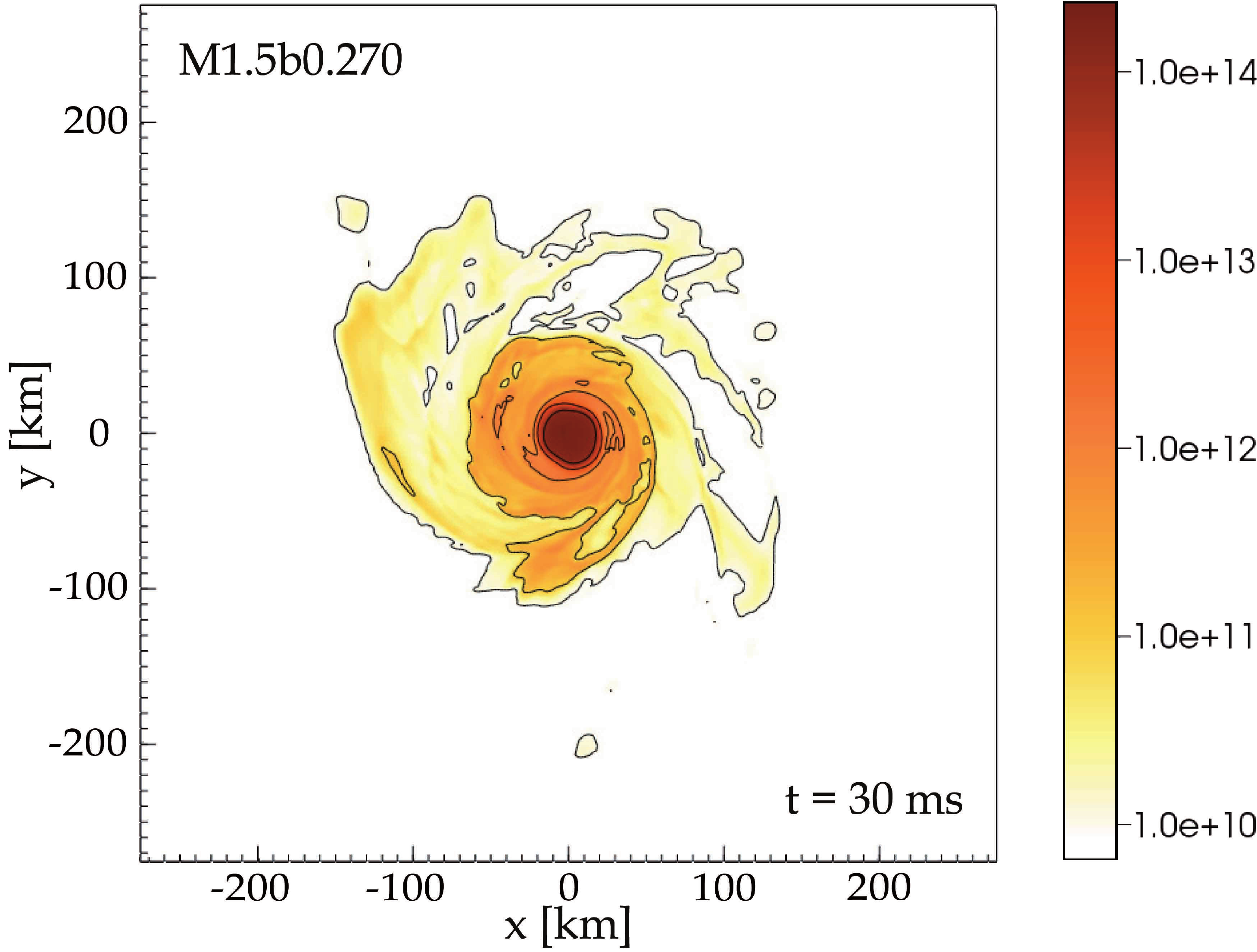} %% 30 ms
\\
\end{tabular}
\end{centering}
\caption{Snapshots of the rest-mass density $\rho$ in the $(x,y)$
  plane for M1.5b0.270 at different stage of the evolution,
  namely $t$=8 and 10 ms (top row), $t$=12 and 14 ms (central row) and
  $t$=20 and 30 ms (bottom row).
The color code is defined in terms of g/cm$^{3}$.
\label{fig:snap_shot}}
\end{figure*}

\begin{figure*}
\vspace{-4mm}
\begin{center}
  \includegraphics[width=0.90\textwidth]{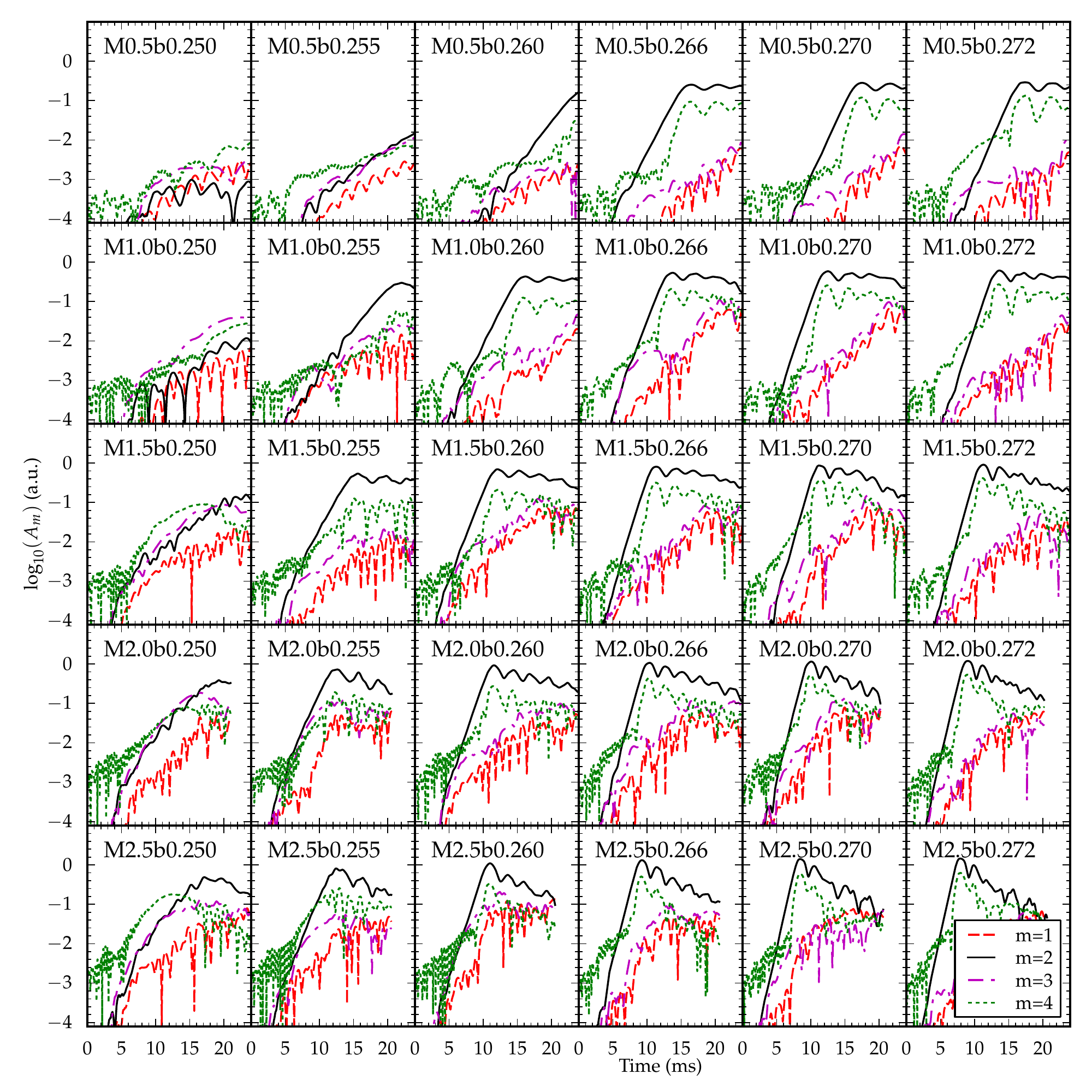}
\end{center}
\vspace{-8mm}
 \caption{Mode dynamics for selected evolved models that are
   characterized by a value of the instability parameter $\beta$
   between 0.250 and 0.272. All models with $\beta \ge 0.255$
   show the typical dynamics one would expect when the dynamical $m=2$ bar
   mode instability is the dominating phenomenon.
\label{fig:high_beta}}
\end{figure*}

\begin{figure*}
\vspace{-4mm}
\begin{center}
  \includegraphics[width=0.90\textwidth]{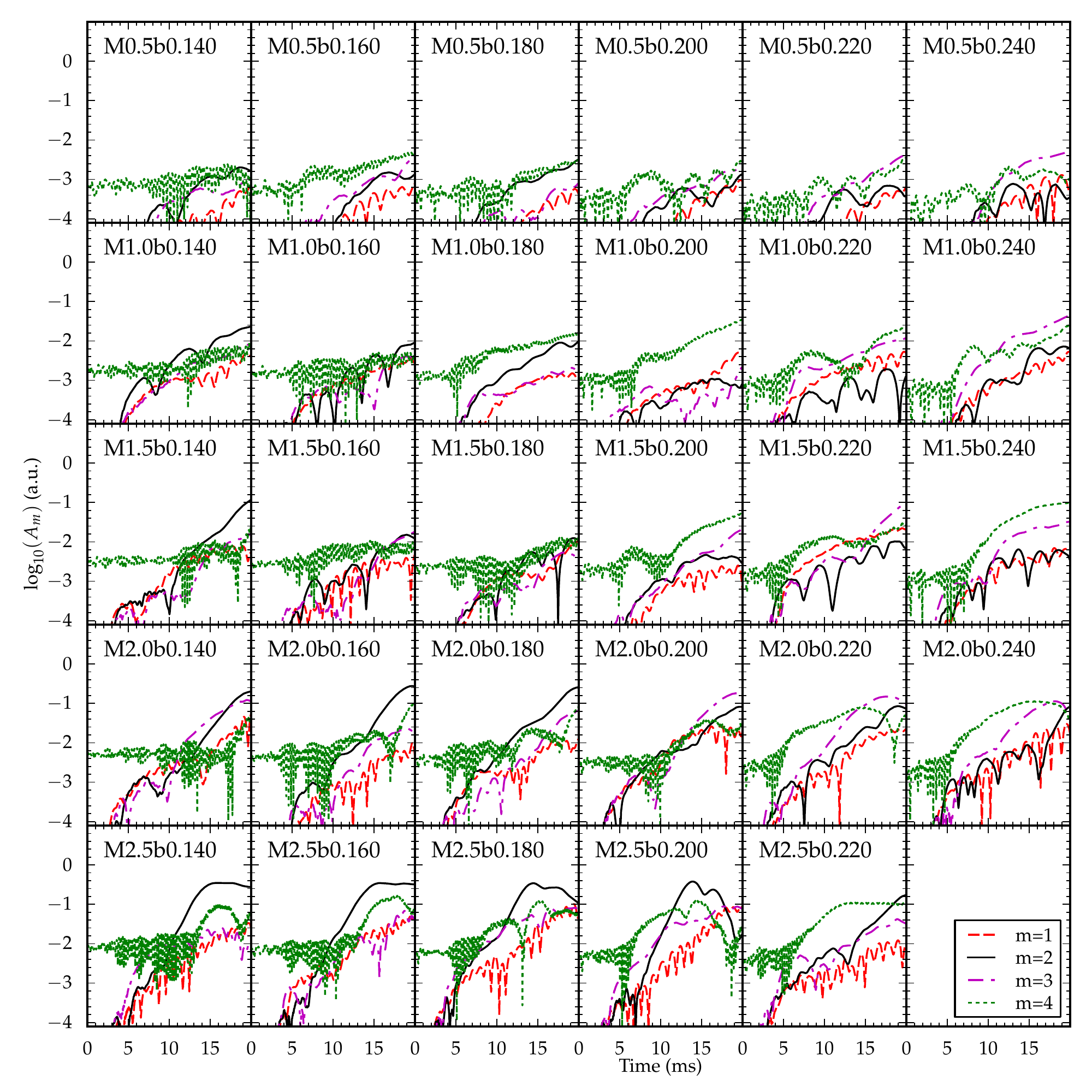}
\end{center}
\vspace{-4mm}
 \caption{The same as Fig.~\ref{fig:high_beta} but for models
   that are characterized by a value of the $\beta$ parameter in the
   range $0.14\le \beta\le 0.24$.
\label{fig:lower_beta}}
\end{figure*}

\subsection{Numerical setup and evolution method}
\label{sec:numsetup}

The main core of the code used for this work is the Einstein
Toolkit~\cite{Loffler:2011ay,EinsteinToolkit:web}, which is a free,
publicly available, community-driven general relativistic (GR) code,
capable of performing numerical relativity simulations that include
realistic physical treatments of matter, electromagnetic
fields~\cite{Moesta:2013dna}, and gravity.

The Einstein Toolkit is built upon several open-source components that
are widely used throughout the numerical relativity community. Among
all of them, only the ones used in this work are mentioned below.

Most components are part of the final evolution code, while others
help managing the components~\cite{Seidel:2010aa,Seidel:2010bb},
building the code and submitting the simulations on
supercomputers~\cite{Thomas:2010aa,SimFactory:web}, or providing
remote debuggers~\cite{Korobkin:2011tg} and post-processing and
visualization tools for \codename{VisIt}~\cite{Childs:2005ACS}.

Many components of the Einstein Toolkit use the \codename{Cactus}
Computational
Toolkit~\cite{Cactuscode:web,Goodale:2002a,CactusUsersGuide:web}, a
software framework for high-performance computing (HPC\@).
\codename{Cactus} simplifies designing codes in a modular
(``component-based'') manner, and many existing \codename{Cactus}
modules provide infrastructure facilities or basic numerical
algorithms such as coordinates, boundary conditions, interpolators,
reduction operators, or efficient I/O in different data formats.

Many of the details of the Einstein Toolkit may be found
in~\cite{Loffler:2011ay}, which describes the routines used to provide
the supporting computational infrastructure for grid setup and
parallelization, constructing initial data, evolving dynamical GRHD
configurations, and analyzing the resulting data describing the
properties of the objects being simulated as well as their
gravitational wave signatures.  The option to evolve magnetic fields,
as described in~\cite{Moesta:2013dna}, is not used here: magnetic
fields are not considered in this work.  In the following, we only
mention or briefly describe the specific methods used for this work
together with the chosen, relevant parameters.  For details the reader
is referred especially to~\cite{Loffler:2011ay}.

Within this study, the adaptive mesh refinement (AMR) methods
implemented by \codename{Carpet}~\cite{Schnetter:2003rb,
  Schnetter:2006pg, CarpetCode:web} have been used.  Carpet supports
Berger-Oliger style (``block-structured'') AMR\@~\cite{Berger:1984zza}
with sub-cycling in time as well as certain additional features
commonly used in numerical relativity (see~\cite{Schnetter:2003rb} for
details).  \codename{Carpet} supports both vertex-centered and
cell-centered AMR\@, but only vertex-centered grids have been used
here.

Hydrodynamic evolution techniques are provided in the Einstein Toolkit
by the \GRHydro package~\cite{Baiotti:2004wn,Hawke:2005zw}.  The code
is designed to be modular, interacting with the vacuum metric
evolution only by contributions to the stress-energy tensor and by the
local values of the metric components and extrinsic curvature, as we
discuss in detail below.  It uses a high-resolution shock capturing
finite-volume scheme.

The evolution of the spacetime metric in the Einstein Toolkit is
handled by the \codename{McLachlan}
package~\cite{McLachlan:web}.  This code is
auto-generated by Mathematica using
\codename{Kranc}~\cite{Husa:2004ip,Lechner:2004cs,Kranc:web},
implementing the Einstein equations via a $3+1$-dimensional split
using the BSSN
formalism~\cite{Nakamura:1987zz,Shibata:1995we,Baumgarte:1998te,
  Alcubierre:2000xu,Alcubierre:2002kk}.

The BSSN equations are finite-differenced at a user-specified order of
accuracy, and coupling to hydrodynamic variables is included via the
stress-energy tensor.  The time integration and coupling with
curvature are carried out with the Method of Lines
(MoL)~\cite{Hyman:1976cm}, implemented in the \codename{MoL} package.
Within this paper a fourth-order
Runge-Kutta~\cite{Runge:1895aa,Kutta:1901aa} method was used, and
Kreiss-Oliger dissipation was applied to the evolved quantities of the
curvature evolution to damp high-frequency noise.

We use fourth-order difference stencils for the curvature evolution,
and $1+\log$~\cite{Alcubierre:2002kk} slicing
\begin{equation}
 \partial_t\alpha = -2\alpha K,
\end{equation}
and $\Gamma$-driver shift condition~\cite{Alcubierre:2002kk},
\begin{align}
 \partial_t\beta^i &= \frac{3}{4} B^i\\ \partial_t B^i &=
 \partial_t\hat\Gamma^i-\frac{1}{2}B^i,
\end{align}
with $K$, $\hat\Gamma^i$, $\alpha$ and $\beta^i$ being the trace of
the extrinsic curvature, the conformal connection functions, the lapse
factor and the shift, respectively.  During time evolution, a
Sommerfeld-type radiative boundary condition is applied to all
components of the evolved BSSN variables as described
in~\cite{Alcubierre:2000xu}.

All presented results use the Marquina Riemann
solver~\cite{Donat:1996cs,Aloy:1999ne} and PPM (the piecewise
parabolic reconstruction method)~\cite{Colella:1982ee}.  An artificial
low-density atmosphere with $\rho_{\text{atm}}=10^{-10}$ is used, with
a threshold of $\rho_{\text{atm\_reset}}=10^{-7}$ below which regions
are reset to atmosphere.  Hydrodynamical quantities are set to
atmosphere at the outer boundary.

All presented evolutions use a mirror symmetry across the $(x,y)$
plane, consistent with the symmetry of the problem, which reduces the
computational cost by a factor of $2$.
In addition, one has the possibility to reduce the computational cost
by an additional factor of $2$, imposing a rotational $\pi$-symmetry
that corresponds to the assumption that the configuration is the same if one
applies a rotation of an angle $\pi$ around the $z$-axis.  However, in
contrast to the mirror symmetry, the $\pi$-symmetry needs to be
justified, because previous
results~\cite{DePietri06,Baiotti:2006wn,Manca:2007ca} showed that
introducing this numerical symmetry by construction prevents odd modes
to grow, something that does in fact happen 
when $\pi$-symmetry is not imposed. Indeed, since we are 
also interested to investigate whether odd modes play any role, we present 
here only results obtained by not imposing $\pi$-symmetry and 
thus we have not taken advance of the 2-fold speedup that 
would have allowed. Please note that results using $\pi$-symmetry 
were presented in~\cite{2013APS..APRY14009L} but they related only to the initial
stage of the evolution of the dynamical bar-mode in the unstable 
region (see SubSec. \ref{subsec:below}) and the results obtained there
have been validated only by the present work.

\section{Results}
\label{sec:results}

As discussed in Sect.~\ref{sec:intro} and~\ref{sec:setup}, the main 
goal of the present work is to study the matter instability that may develop
in the case of rapidly differentially rotating relativistic star models,
using a configuration as close as possible to the realistic case.
The other important requirement we need to fulfill is that our study has to be 
computationally feasible. To achieve this goal, we need to evolve the largest 
number of models using the available amount of computational resources 
in the most efficient way. 
In selecting a numerical setting we can play with many 
parameters, namely: the location of the outer boundary, the number 
of refinement levels, the size and resolution of the finest grid
and the symmetries to be imposed on the dynamics. 
All the simulations in the present work are performed 
using the same setting for the computational domain. More precisely, we use 
three box-in-box (covering the half space with $z\ge0$) refinement levels, 
with boundaries at distances of $L=42,84,168$ from the 
origin of the coordinate system and grid spacings $dx,2\,dx,4\,dx$, respectively,
where we set $dx=0.5$ (that correspond to a resolution $dx\simeq 0.738$ km) 
unless otherwise noted, corresponding to a hierarchy of 
three computational grids, each one of size $169\times169\times85$ 
points plus ghost and buffer zones.

We have chosen to use this conservative domain, large enough to capture the whole global 
dynamics of a bar-mode instability, in order to exclude any influence of the computational setup 
on observed differences between models. The actual size of the finest grid and the 
computational set up is determined by the most demanding models.
Fig.~\ref{fig:snap_shot} shows a few snapshots for the 
evolution of the rest-mass density $\rho$ at different times for
a representative model, namely M1.5b0.270 which is characterized by
$\beta=0.270$ and $M_0=1.5 \,M_\odot$. This is indeed the typical
evolution one would expect for a stellar model which is unstable against 
the dynamical bar-mode instability.

\begin{figure}
\vspace{-1mm}
 \begin{center}
  \includegraphics[width=0.45\textwidth]{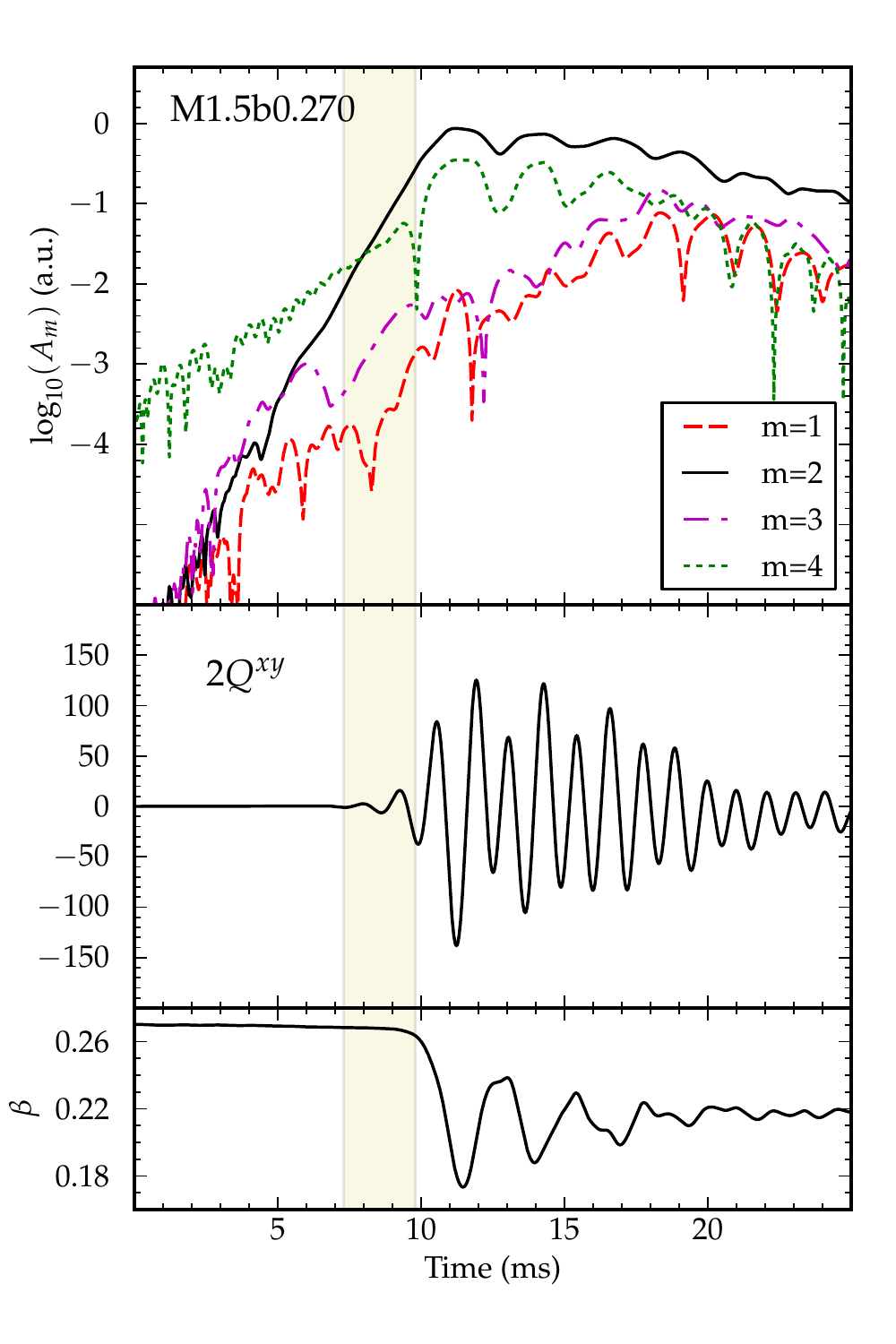}
 \end{center}
\vspace{-8mm}
 \caption{Main features of the dynamics of a representative example of
   an evolved stellar model that is unstable against the bar-mode
   instability, namely M1.5b0.270. In particular, we show the time evolution of the 
   power of the cylindrical $m=1,\,2,\,3,\,4$ modes of the matter density (upper panel),
   the time evolution of the $xy$ component of the quadrupole moment of the conserved
   density $Q^{xy}$, defined in Eq.~(\ref{eq:defQxy}) (center panel), and the time evolution of the 
   rotational parameter $\beta=T/|W|$. In the shaded region, corresponding to 
   $7$ ms  < $t$ < $10$ ms,
   there is a clear exponential growth of the $m=2$ azimuthal matter
   mode that later reaches a saturation at 
   $t \approx 12$ ms. Eventually, the model seems to settle down around a new
   (less differentially rotating) 
   configuration after $t \approx 20$ ms.
\label{fig:evolution_example}}
\end{figure}

\subsection{Analysis Methods}
\label{sec:analysis}

In order to compute the growth time of the instability, $\tau_2$, we use the
quadrupole moments of the matter distribution $Q^{ij}$, computed in terms of 
the conserved density $D$ as
\begin{equation}
\label{eq:defQxy}
Q^{ij} = \int\! d^{3}\!x \; D \; x^{i} x^{j} \ .
\end{equation}
In particular, we perform a nonlinear least-square fit of $Q^{xy}$
 (the star spin axis is aligned in the $z$-direction), using the trial function
\begin{equation}
 \label{eq:Qxy}
 Q^{xy}(t) = Q^{xy}_0\mathrm{e}^{\frac{t}{\tau_2}}\cos(2\pi f_2 t+\phi_0) \, .
\end{equation}
Using this trial function, we can extract the growth time $\tau_2$ and the frequency
$f_2$ for the unstable $m=2$ modes.
We also define the modulus $Q(t)$ as
\begin{equation}
 \label{eq:Q}
 Q \equiv\frac{1}{2}\sqrt{(2Q^{xy})^2 + (Q^{xx}-Q^{yy})^2} \, ,
\end{equation}
and the distortion parameter $\eta(t)$ as
\begin{equation}
 \label{eq:def_eta}
 \eta \equiv \frac{2 Q}{(Q^{xx}+Q^{yy})} \, .
\end{equation}
Finally, we decompose the rest-mass density into its spatial
rotating modes $P_m(t)$
\begin{equation}
 P_m \equiv \int\diff^3x \, \rho \mathrm{e}^{\mathrm{i}m\phi} 
\end{equation}
and the ``amplitude'' and ``phase'' of the $m$-th mode are defined as
\begin{equation}
 \label{eq:Am}
A_m = |P_m| \,\quad\, \text{and} \,\quad\, \phi_m \equiv arg(P_m) \, .
\end{equation}
Despite their denomination, the amplitudes defined in Eq.~(\ref{eq:Am}) do not
correspond to proper eigenmodes of oscillation of the star but to global
characteristics that are selected in terms of their spatial azimuthal shape.  
All Eqs.~(\ref{eq:Qxy})-(\ref{eq:Am}) are expressed in terms of the
coordinate time $t$, and therefore they are not gauge invariant. However,
the length scale of variation of the lapse function at any given time
is always small when compared to the stellar radius, ensuring that events
close in coordinate time are also close in proper time.

\subsection{General features of the evolution above the
threshold for the onset of the bar-mode instability}
\label{subsec:above}

The above mentioned general features of the evolution are common to
all the modes that show the expected dynamics in presence of the well
studied bar-mode $m=2$ instability. In Fig.~\ref{fig:high_beta} the
``mode-dynamics'' of most of the studied models having $\beta\ge 0.25$
are shown. For all these models (except for M0.5b0.250 and M1.0b0.250)
it is indeed possible to extract the main features of the $m=2$ mode
using the procedure detailed in Eq.~(\ref{eq:Qxy}).
Fig.~\ref{fig:evolution_example} shows the time
evolution of some quantities that characterize the behavior of model
M1.5b0.270, for which also some snapshots were shown in
Fig.~\ref{fig:snap_shot}.  We decided to quantify the properties of
the bar-mode instability by means of a nonlinear fit, using
the trial dependence of Eq.~(\ref{eq:Qxy}) on a time interval where the
distortion parameter $\eta$ defined in Eq.~(\ref{eq:def_eta}) is
 between $1\%$ and $30\%$ of its maximum value. The
shaded region in Fig.~\ref{fig:evolution_example} corresponds exactly 
to the region we selected for the fit according to this criterion.

\begin{figure}
\vspace{-1mm}
 \begin{centering}
  \includegraphics[width=0.45\textwidth]{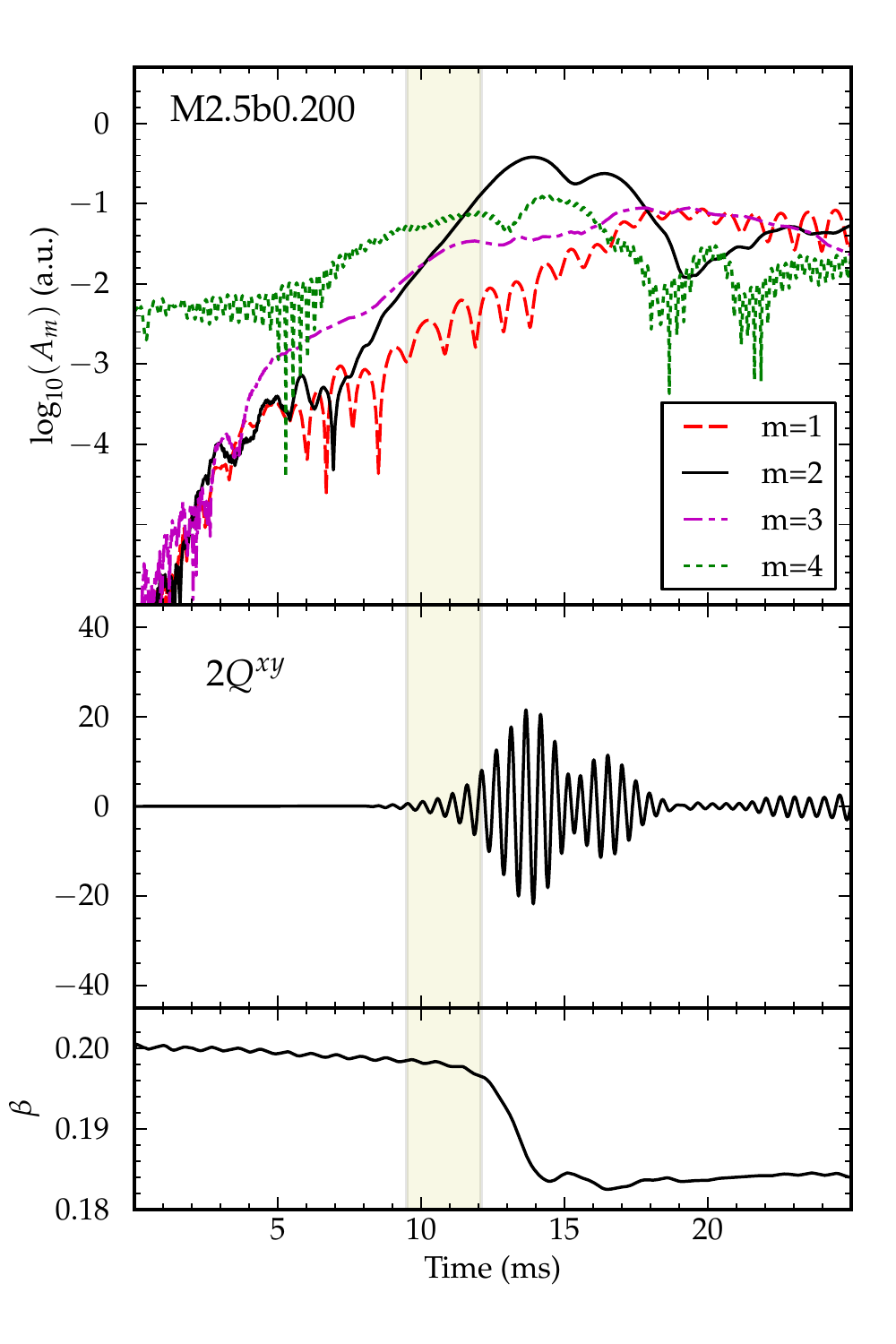}
 \end{centering}
\vspace{-4mm}
\caption{The same as Fig.~\ref{fig:high_beta} but for model
  M2.5b0.200, that shows the presence of a $m=2$ shear instability at
  a low value of $\beta$. The dynamics enclosed in the shaded region
  has been selected to compute the characteristic frequency and growth
  time of the $m=2$ instability, that turned out to be $f_2=1.943$ kHz
  and $\tau_2=1.02$ ms respectively.
\label{fig:evolution_shear}}
\end{figure}

The results of all these fits are collected in
Tab.~\ref{tab:res_dx05}, where we report for each model the 
maximum value assumed by the distortion parameter $\max(\eta)$, the time
interval $[t_i,t_f]$ selected for the fit, the value
$\beta(t_i)$ corresponding to the value of the instability parameter
$\beta$ at the beginning of the fit interval and $\tau_2$ and $f_2$,
the growth time and frequency that characterize the $m=2$ bar-mode
instability, respectively.

\begin{figure}
\vspace{-1mm}
 \begin{centering}
  \includegraphics[width=0.45\textwidth]{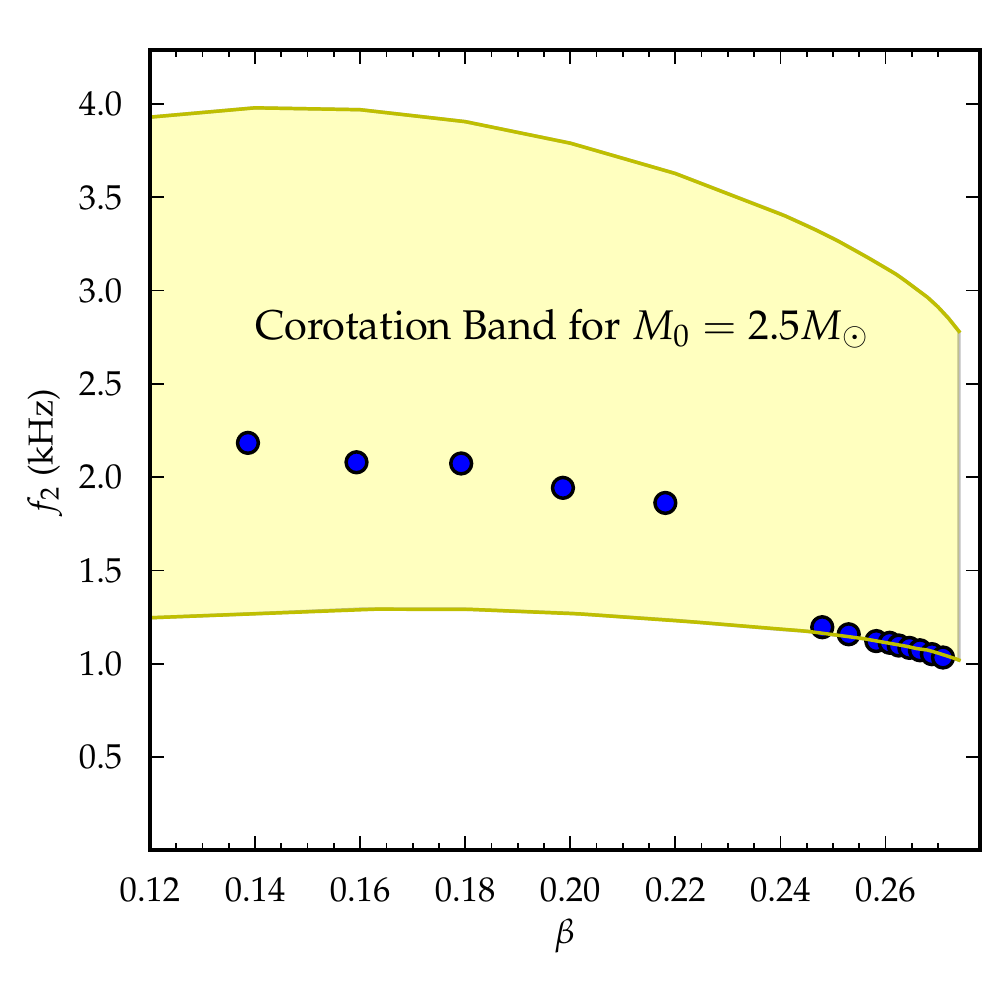}
 \end{centering}
\vspace{-4mm}
 \caption{Here we show the frequency of the corotation band multiplied
   by 2 (shaded region) and the measured values for the coordinate
   frequency of the observed $m=2$ modes (circles) for all the
   models belonging to the sequence with baryon mass $M_0=2.5 M_\odot$.
\label{fig:freqM25}}
\end{figure}

\subsection{General features of the evolution below the
threshold for the onset of the bar-mode instability}
\label{subsec:below}

The situation is more complicated for initial models characterized by
lower values of the $\beta$ parameter, i.e. $\beta < 0.245$.  For
these models (see Fig.~\ref{fig:lower_beta}) one can observe
there is an indication that instabilities are present. 
For example, models like M0.5b0.200, M0.5b0.220,
M0.5b0.240 and M2.0b0.200, M2.0b0.220, M2.0b0.240 show the possible
presence of a three-arms, $m=3$, unstable mode. Other models, e.g. 
M2.0b0.140, seem to show a competition between two different 
unstable modes, namely the $m=2$ mode and the $m=3$ mode. 

Other models show the presence of $m=2$ unstable modes. We use
as a practical criteria to select such models the fact that they have 
a maximal distortion parameter $\eta$ greater than $10\%$, i.e. $\max(\eta) > 0.10$.  
The simulated models that fulfill
this criteria are: 
models M1.0b0.140 and M1.0b0.160, belonging to the sequence with $M_0=1.0$;
model M1.5b0.140 for $M_0=1.5$;
models M2.0b0.140, M2.0b0.160, M2.0b0.180 for $M_0=2.0$ and, 
finally, models M2.5b0.140, M2.5b0.160, M2.5b0.180, M2.5b0.200 and M2.5b0.220
for the $M_0=2.5$ sequence.
For example, in the case of model M2.5b0.200 (see Fig.~\ref{fig:evolution_shear}) 
we can observe an exponential growth,
with a frequency $f_2=1.95$ kHz and a growth time $\tau_2 \simeq 1$,
of the $m=2$ mode developing and eventually saturating at about 12 ms,
when the model settles down to a new equilibrium configuration
corresponding to a lower value of the rotational parameter $\beta$ and
a different differential rotation profile. Indeed, the frequency of all
these $m=2$ modes are inside the corotation band (see Fig.~\ref{fig:freqM25}) 
and should be classified as shear instabilities, of the same type of those
observed in~\cite{Corvino:2010}. 

The same check has been performed for all unstable modes (both those
with $m=2$ and those with $m=3$) that are indeed shear instabilities.
In particular, for the $m=2$ of the models with $M_0=2.5$ below the 
threshold for the onset of the classical dynamical 
bar-mode instability, in Fig.~\ref{fig:freqM25} it is shown that 
the frequency $f_2$ of the unstable model are within the corotation
band.

\begin{table}
\vspace{-1mm}
\begin{center}\begin{tabular}{|l|c|rr|rcc|}
\hline model & $\max(\eta)$ & $t_i$ & $t_f$ & $\beta(t_i)$ & $\tau_2$(ms) & $f_2$(kHz)\\
\hline
\hline
M0.5b0.255 & 0.178 & 13.2 & 26.2 & 0.2527 & 3.913 & 0.527 \\ 
M0.5b0.260 & 0.404 & 14.8 & 21.3 & 0.2573 & 1.899 & 0.515 \\ 
M0.5b0.262 & 0.473 & 12.3 & 17.7 & 0.2597 & 1.604 & 0.512 \\ 
M0.5b0.264 & 0.515 & 13.4 & 18.4 & 0.2612 & 1.474 & 0.508 \\ 
M0.5b0.266 & 0.578 &  9.2 & 13.7 & 0.2639 & 1.307 & 0.505 \\ 
M0.5b0.268 & 0.615 & 11.2 & 15.4 & 0.2656 & 1.223 & 0.502 \\ 
M0.5b0.270 & 0.664 & 11.1 & 15.0 & 0.2674 & 1.150 & 0.496 \\ 
M0.5b0.272 & 0.713 & 11.3 & 15.0 & 0.2695 & 1.085 & 0.493 \\ 
\hline
\hline
M1.0b0.255 & 0.475 & 11.2 & 18.0 & 0.2532 & 1.959 & 0.685 \\ 
M1.0b0.260 & 0.702 &  9.2 & 13.5 & 0.2584 & 1.256 & 0.673 \\ 
M1.0b0.262 & 0.776 &  8.4 & 12.3 & 0.2605 & 1.152 & 0.668 \\ 
M1.0b0.264 & 0.836 &  8.6 & 12.2 & 0.2623 & 1.037 & 0.663 \\ 
M1.0b0.266 & 0.893 &  8.3 & 11.6 & 0.2645 & 0.964 & 0.658 \\ 
M1.0b0.268 & 0.936 &  8.3 & 11.5 & 0.2665 & 0.908 & 0.652 \\ 
M1.0b0.270 & 0.992 &  7.5 & 10.5 & 0.2685 & 0.863 & 0.646 \\ 
M1.0b0.272 & 1.021 &  8.9 & 11.8 & 0.2698 & 0.826 & 0.639 \\ 
\hline
\hline
M1.5b0.250 & 0.180 &  6.7 & 17.1 & 0.2494 & 3.260 & 0.860 \\ 
M1.5b0.255 & 0.658 &  8.1 & 12.8 & 0.2537 & 1.380 & 0.835 \\ 
M1.5b0.260 & 0.864 &  6.7 & 10.1 & 0.2589 & 0.976 & 0.816 \\ 
M1.5b0.262 & 0.908 &  8.5 & 11.7 & 0.2604 & 0.949 & 0.809 \\ 
M1.5b0.264 & 0.974 &  7.5 & 10.4 & 0.2624 & 0.853 & 0.802 \\ 
M1.5b0.266 & 1.043 &  6.9 &  9.6 & 0.2648 & 0.789 & 0.796 \\ 
M1.5b0.268 & 1.086 &  7.6 & 10.2 & 0.2666 & 0.747 & 0.787 \\ 
M1.5b0.270 & 1.123 &  7.3 &  9.8 & 0.2683 & 0.721 & 0.778 \\ 
M1.5b0.272 & 1.175 &  7.3 &  9.7 & 0.2699 & 0.696 & 0.770 \\ 
\hline
\hline
M2.0b0.250 & 0.362 &  7.6 & 14.8 & 0.2486 & 2.203 & 1.023 \\ 
M2.0b0.255 & 0.749 &  6.4 & 10.2 & 0.2536 & 1.140 & 0.988 \\ 
M2.0b0.260 & 0.917 &  6.9 &  9.8 & 0.2582 & 0.849 & 0.969 \\ 
M2.0b0.262 & 0.995 &  7.3 & 10.0 & 0.2604 & 0.806 & 0.953 \\ 
M2.0b0.264 & 1.059 &  7.2 &  9.7 & 0.2628 & 0.731 & 0.942 \\ 
M2.0b0.266 & 1.121 &  6.5 &  8.8 & 0.2647 & 0.687 & 0.934 \\ 
M2.0b0.268 & 1.155 &  6.6 &  8.9 & 0.2667 & 0.650 & 0.923 \\ 
M2.0b0.270 & 1.203 &  6.5 &  8.6 & 0.2686 & 0.626 & 0.912 \\ 
M2.0b0.272 & 1.245 &  5.6 &  7.6 & 0.2707 & 0.596 & 0.900 \\ 
\hline
\hline
M2.5b0.250 & 0.372 &  7.1 & 13.7 & 0.2480 & 1.843 & 1.195 \\ 
M2.5b0.255 & 0.684 &  7.0 & 10.6 & 0.2530 & 1.042 & 1.158 \\ 
M2.5b0.260 & 0.922 &  6.7 &  9.4 & 0.2583 & 0.798 & 1.121 \\ 
M2.5b0.262 & 1.010 &  5.7 &  8.1 & 0.2608 & 0.711 & 1.112 \\ 
M2.5b0.264 & 1.073 &  5.6 &  7.9 & 0.2625 & 0.667 & 1.097 \\ 
M2.5b0.266 & 1.118 &  5.8 &  7.9 & 0.2646 & 0.627 & 1.085 \\ 
M2.5b0.268 & 1.173 &  5.5 &  7.5 & 0.2666 & 0.586 & 1.072 \\ 
M2.5b0.270 & 1.221 &  5.3 &  7.3 & 0.2688 & 0.565 & 1.051 \\ 
M2.5b0.272 & 1.261 &  4.9 &  6.8 & 0.2710 & 0.541 & 1.033 \\ 
\hline
\end{tabular}\end{center}
\vspace{-2mm}
\caption{Results for the growth time $\tau_2$ and the frequency $f_2$
  of the bar-mode for all the models that show the presence of the
  bar-mode instability. The values are obtained from simulations 
  which employ a spatial resolution $dx=0.5$ for the finest grid.}
\label{tab:res_dx05}
\end{table}

\subsection{Effects of the compactness on the threshold for the onset of the bar-mode 
instability}
\label{subsec:BAR}

We have chosen to investigate the effect of the compactness on the
classical bar-mode instability, at fixed stiffness, following the same
procedure as in~\cite{Manca:2007ca}. We determined the critical value
of the instability parameter $\beta$ for the onset of the instability
by simulating five sequences of initial models having the same value of
$M_0$ but different values of $\beta$. For these simulations we
decided to employ the same resolution $dx=0.5$ on the finest grid 
for all the simulations. This choice was mainly motivated by the necessity 
to keep the computational cost under reasonable limits.

We now restrict our analysis to the models for which we observed the maximum
value of the distortion parameter $\eta$ to be greater than $0.10$. 
For these models, we explicitly checked that the reported unstable
modes correspond to the classical bar-mode instability and not to a
shear-instability by checking that the frequency of the mode divided by
two is not in the corotation band of the model (see
Fig.~\ref{fig:corotation_band}). 
This is effectively true for all the reported models, except for M2.5b0.250, 
M2.5b0.255 and M2.5b0.260, which are just marginally (at the lower boundary) 
on the corotation band (see Fig.~\ref{fig:freqM25}).

\begin{figure}
\vspace{-1mm}
\begin{centering}
\includegraphics[width=0.45\textwidth]{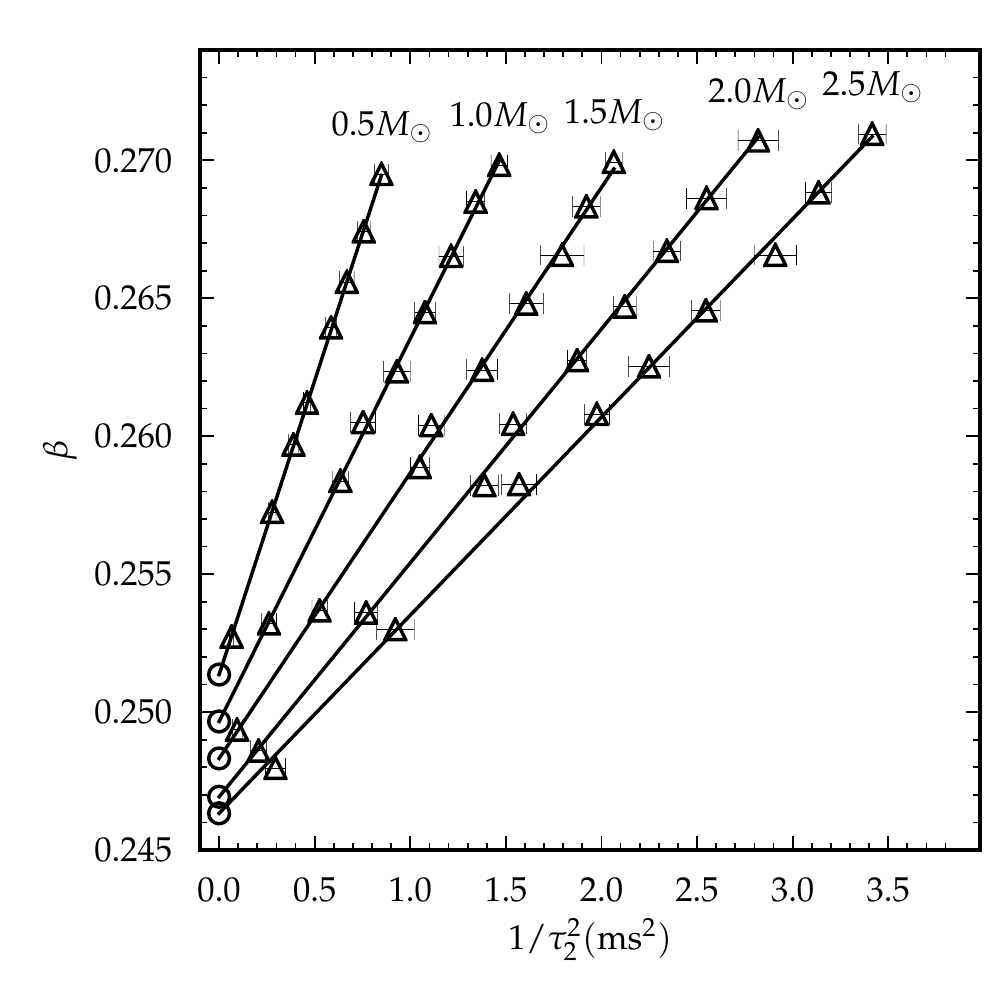}
\end{centering}
\vspace{-4mm}
 \caption{Critical diagram relating the growth time $\tau_2$ of each
   bar-mode unstable model to the value of the instability parameter
   $\beta$. Triangles represent the values corresponding to all the
   models listed in Tab.~\ref{tab:res_dx05}.  More precisely, the
   quantity on the $x$-axis is actually expressed in terms of $1 /
   \tau_2^2$, in order to highlight the very good fit, while the quantity
   on the $y$ axis is the value of $\beta$ at the beginning of the
   time interval chosen for performing the fit of $m=2$ mode growth,
   reported in Tab.~\ref{tab:res_dx05} as $\beta(t_i)$.  For all the
   five constant rest-mass sequences considered we also report, with
   open circles, the extrapolated value $\beta_c$ and the fitted 
   straight lines with Eqs.~(\ref{eq:res_fits}).
\label{fig:growthtimes}}
\end{figure}

\begin{figure}
\begin{centering}
\includegraphics[width=0.45\textwidth]{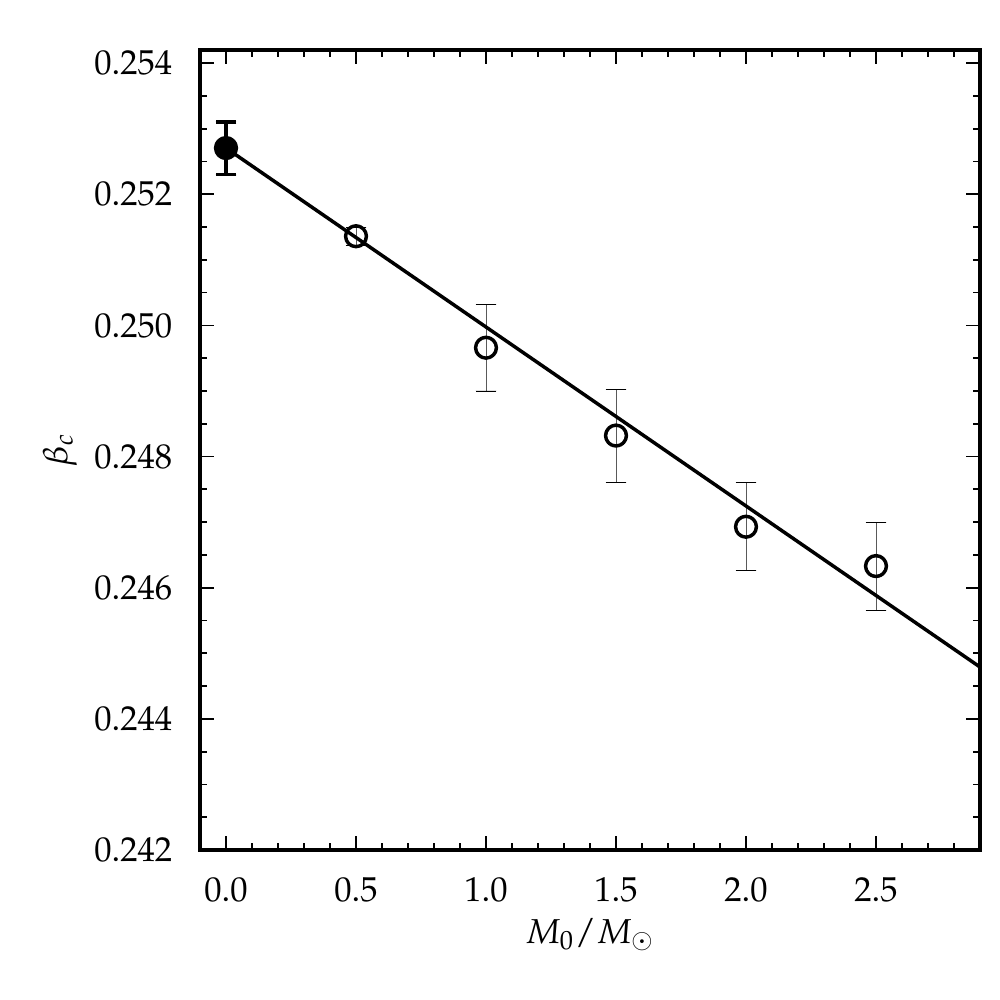}
\end{centering}
 \caption{The open circles are the same as in Fig.~\ref{fig:growthtimes},
   but here they are shown
   as a function of the baryonic mass $M_0$. The filled circle, on the
   other hand, marks the extrapolation for zero rest-mass, representing
   the limit of Newtonian gravity (or zero compactness). The result of
   this extrapolation is reported in Eq.~(\ref{eq:NewtonianExtrapolation}).
\label{fig:NewtonianLimit}}
\end{figure}

We have performed a fit for the growth time $\tau_2$ of the bar mode
as a function of the instability parameter $\beta$ for five
sequences of models with constant rest-mass ranging from 0.5 $M_\odot$
to 2.5 $M_\odot$, as shown in Fig.~\ref{fig:growthtimes}. Let us
estimate the threshold for the onset of the instability using the
extrapolation technique used in~\cite{Manca:2007ca} where we assume,
in analogy with what expected in the Newtonian case, that the main
dependence of the frequency of the mode on $\beta$ is of the type
\begin{equation}
 \sigma(\beta) = \Omega(\beta) \pm \sqrt{F(\beta)}\,,
\end{equation} 
where
\begin{equation}
 \frac{1}{(\tau_2(\beta))^2} = F(\beta) \simeq A (\beta_c-\beta)\, .
\end{equation} 
Under this assumption, we find the following values for the critical
fit of the growth times:
\begin{equation}
\begin{aligned}
0.5 M_\odot &  :  &  1/{(\tau_2)^2}&=  6.85(4)  &\!\!\!\!  \times \, ( 0.2512(2) - \beta )\\
1.0 M_\odot &  :  &  1/{(\tau_2)^2}&=  8.46(20) &\!\!\!\!  \times \, ( 0.2497(7) - \beta )\\
1.5 M_\odot &  :  &  1/{(\tau_2)^2}&=  9.83(24) &\!\!\!\!  \times \, ( 0.2483(7) - \beta )\\
2.0 M_\odot &  :  &  1/{(\tau_2)^2}&= 10.86(23) &\!\!\!\!  \times \, ( 0.2469(7) - \beta )\\
2.5 M_\odot &  :  &  1/{(\tau_2)^2}&= 11.80(35) &\!\!\!\!  \times \, ( 0.2463(7) - \beta )
\end{aligned}
\label{eq:res_fits}
\end{equation}

The results obtained so far cannot be directly compared with those
obtained in~\cite{Manca:2007ca} to infer the effects of considering a
stiffer EOS. The main issue is that when considering a polytropic EOS,
one can change the units of measurement in such a way that the value
of the polytropic constant $K$ is $1$. This means that by changing this
value one effectively changes the mass scale and, in turn, the mass of
the considered stellar model.  Indeed, the assertion that for a star
with mass $M_0= 1.0 M_\odot$ the threshold for the onset of the
bar-mode instability is reduced to $0.2497(7)$ for $\Gamma=2.75$ from
the higher value of $0.2598(8)$ for $\Gamma=2.0$ is susceptible to the
choice of the mass scale determined by the choice of the values of the
polytropic constants.  This dependence from the choice of the mass
scale can be eliminated by going to the zero-mass limit that
corresponds to performing an extrapolation to the {\it Newtonian} limit
of the results. This can be achieved by a linear fit of the
reported values for the critical $\beta_c$ for the onset of the
classical bar-mode instability of Eqs.~(\ref{eq:res_fits}) as a function
of the baryonic rest-mass (see Fig.~\ref{fig:NewtonianLimit}). The
overall result for this fit leads to the following expression for the
critical $\beta_c$ as a function of the the total baryonic mass $M_0$:
\begin{align}
\beta_c(M) & = 0.2527(4) - 0.0027(5) M_0 \, .
\label{eq:NewtonianExtrapolation}
\end{align}

The extrapolated value for $\beta_c$ in the
limit of zero baryonic mass for the relativistic stellar models then
leads to
$\beta_c^{N}=0.2527(4)$ for $\Gamma=2.75$, which can now be directly
compared to the one obtained in~\cite{Manca:2007ca}, i.e.
$\beta_c^{N}=0.266(1)$ for $\Gamma=2.0$. A further consistency check
that the extrapolated threshold actually corresponds to the Newtonian
value can be found in the literature for the case of $\Gamma=2$.  In fact,
of the four models discussed in~\cite{Saijo2008,2008PhRvD..78l4001K},
the one characterized by $\beta=0.281,\,0.277,\,0.268$ are unstable,
while the one characterized by $\beta=0.256$ is stable.

This shows that the threshold for the onset of the dynamical bar-mode
instability is significantly but not greatly
reduced by an increase in the stiffness of the EOS, induced
by a change of $\Gamma$ from 2 to 2.75. Unfortunately, this threshold
is very close to the maximum possible value for $\beta$ that can be
sustained by a realistic EOS like the one obtained from the SLy
prescription. As it was shown in~\cite{Corvino:2010}, there are very
few relativistic models with $\hat{A}=1$ that can be generated with a
value of $\beta$ above the threshold for the dynamical bar-mode
instability even if we consider the effect of using a stiffer EOS.

\subsection{Resolution}
\label{subsec:resolution}

\begin{figure}
\vspace{-1mm}
\begin{centering}
  \includegraphics[width=0.48\textwidth]{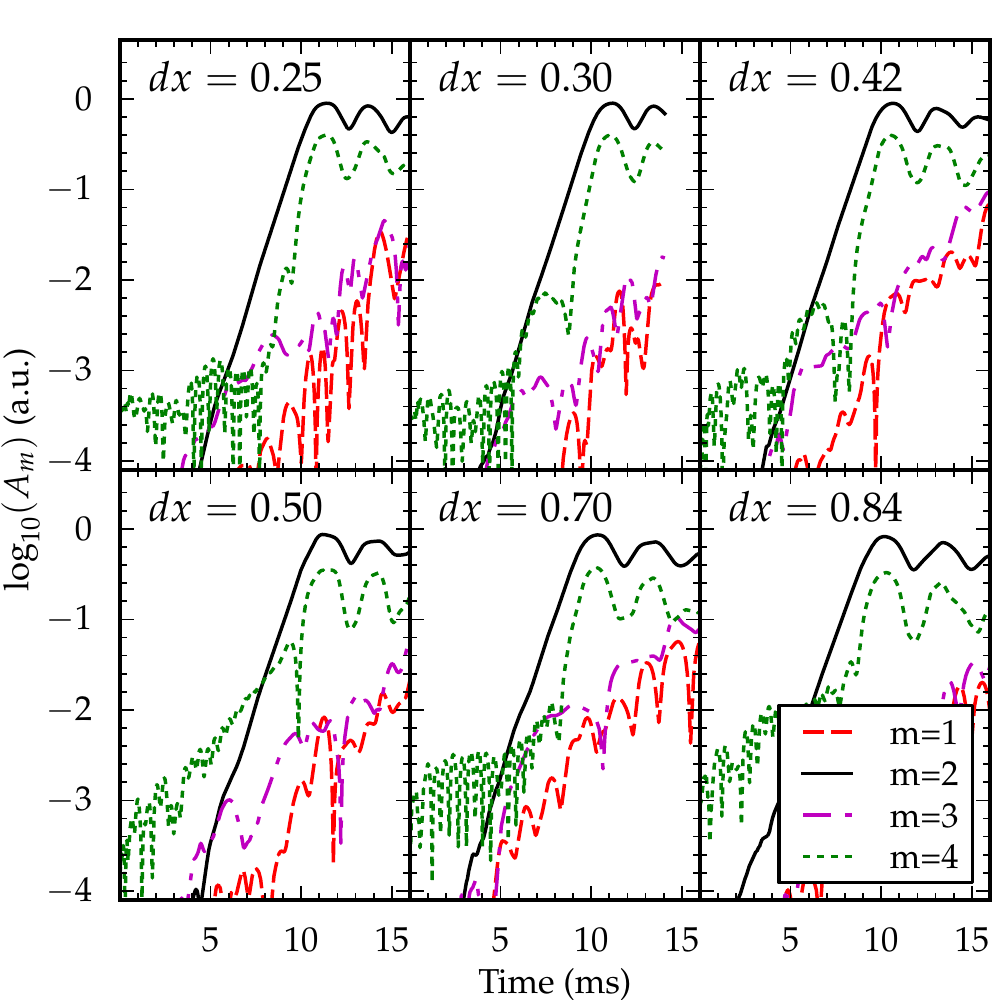}
\end{centering}
\vspace{-4mm}
\caption{Evolution of M1.5b0.270 for different values of the resolution
on the finest grid. 
Note that the total computational cost of the simulation at resolution $dx=0.25$ 
is 16 times greater than a simulation at dx=0.5.  
\label{fig:M15b270res}}
\end{figure}

In order to asses the correctness of the extracted value of $\beta^N_c$ one has to
check how the result depends on the actual resolution
used. To perform this check we have evolved a typical model (M1.5b0.270)
characterized by the same value of the initial baryonic
mass $M_0=1.5 M_\odot$ and a value of the initial rotation parameter
$\beta=0.270$ using different grid resolutions, namely varying the grid
spacing in the range from $dx=0.25$ to $dx=0.84$ in
dimensionless units where $G=c=1$. That corresponds to resolutions 
varying between $dx=0.369$ km and $dx=1.240$ km.
The results of the mode evolutions
we have obtained are shown in Fig.~\ref{fig:M15b270res}. The mode
dynamics we obtained show that the overall picture of the dynamics does
not change.

However the actual results of the fits, as expected, show a
dependence on the used resolution. In Table~\ref{tab:fit_M15b270res} we report:
the resolution used, the maximum value reached by the distortion parameter
$\eta$, the time $t_i$ and $t_f$ for which the distortion parameter has a value
between the 1\% and the 30\% of the maximum, the value of the rotational
parameter at the time $t_i$ and the fitted values for the growth time
and frequency of the unstable bar-mode.

\begin{table}
\begin{center}
\begin{tabular}{|l|c|rr|ccc|}
\hline
resolution & max($\eta$) &  $t_i$ & $t_f$ & $\beta(t_i)$ & $\tau_2$(ms) & $f_2$(kHz)\\
\hline
dx=0.25 (0.369 km)& 1.164 &  7.4 &  9.8 & 0.2697 & 0.692 & 0.783 \\ 
dx=0.30 (0.443 km)& 1.154 &  7.1 &  9.5 & 0.2695 & 0.693 & 0.784 \\ 
dx=0.42 (0.620 km)& 1.147 &  6.6 &  9.0 & 0.2692 & 0.706 & 0.780 \\ 
dx=0.50 (0.738 km)& 1.123 &  7.3 &  9.8 & 0.2683 & 0.721 & 0.778 \\ 
dx=0.70 (1.034 km)& 1.111 &  6.1 &  8.6 & 0.2675 & 0.742 & 0.770 \\ 
dx=0.84 (1.240 km)& 1.049 &  5.9 &  8.5 & 0.2658 & 0.768 & 0.771 \\ 
\hline
\end{tabular}
\end{center}
\vspace{-2mm}
\caption{Results of the fitw of the grow times $\tau_2$ as a function 
of the resolution for model M1.5b0.270.
\label{tab:fit_M15b270res}}
\end{table}

\begin{figure}
\vspace{-1mm}
\begin{centering}
  \includegraphics[width=0.45\textwidth]{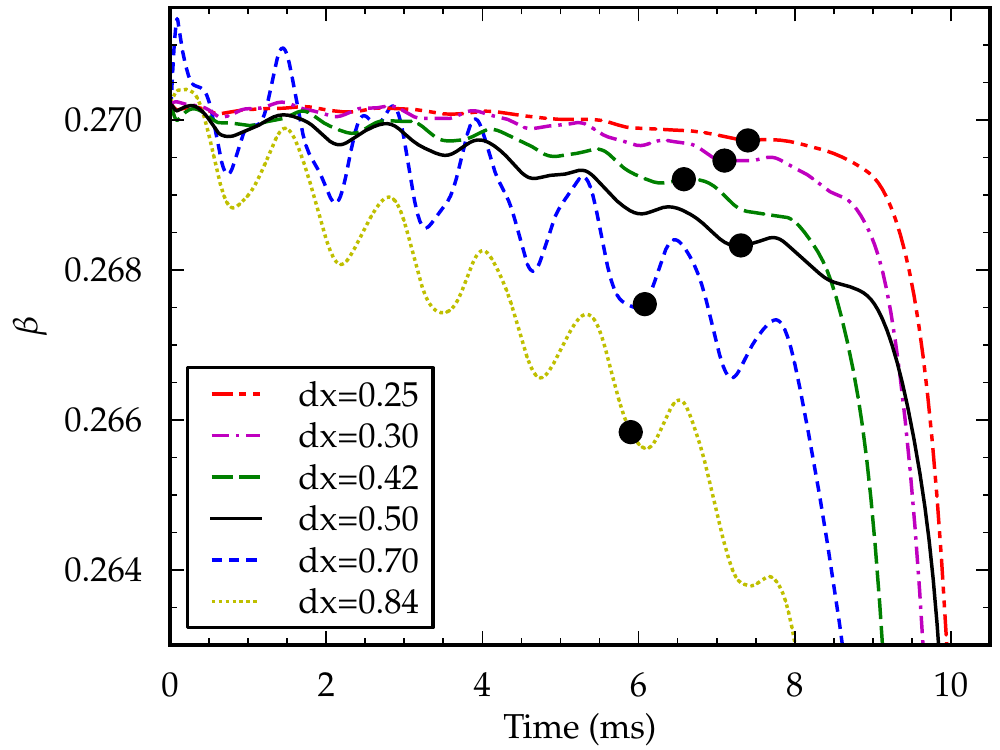}
\end{centering}
\vspace{-4mm}
\caption{Here we show the time evolution of the $\beta$ parameter
  for different values of the employed resolution. 
  The filled dots mark the time at which the values $\beta(t_i)$, used in the fits
  are evaluated.}
\label{fig:beta_evolution_res}
\end{figure}

\begin{figure}
\vspace{-1mm}
\begin{centering}
  \includegraphics[width=0.45\textwidth]{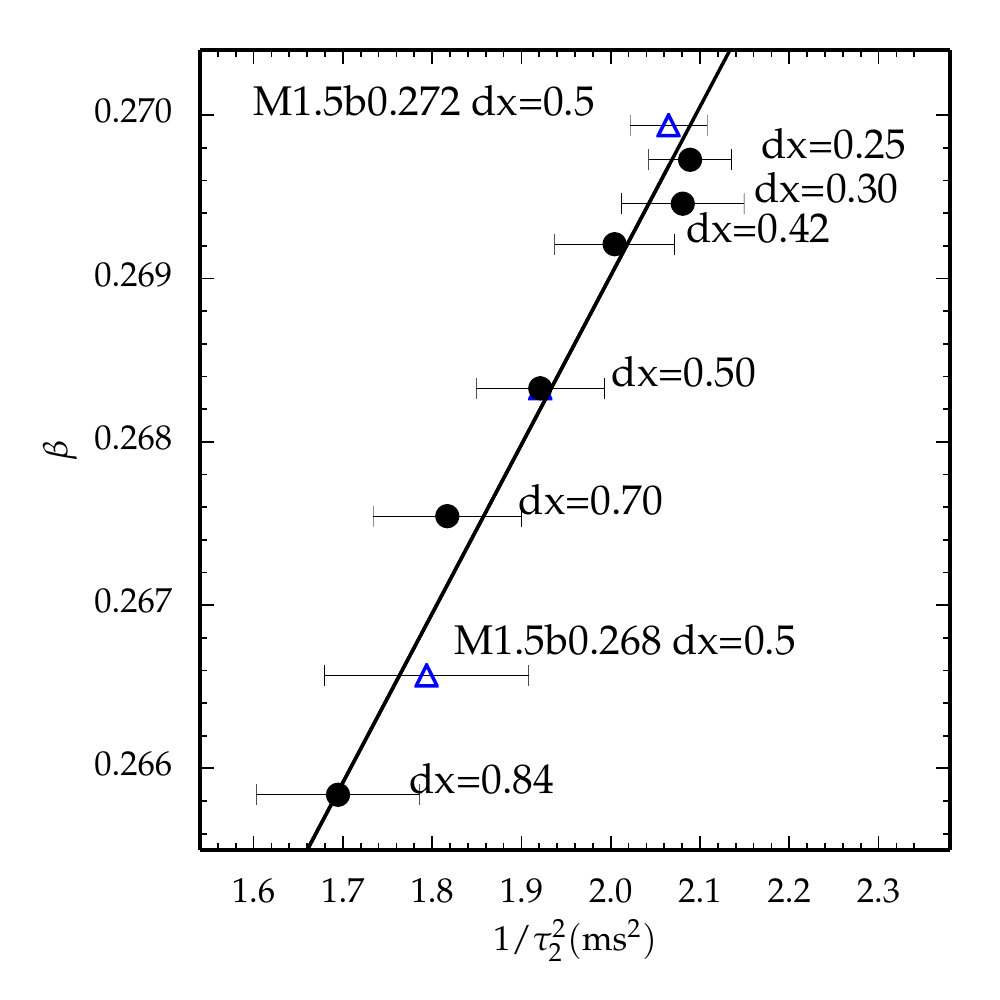}
\end{centering}
\vspace{-4mm}
\caption{The filled circles represent the values of $\beta$ at the
  time when the distortion parameter $\eta$ reaches 1\% of its maximum
  value, as a function of the inverse of the square of the growth
  time, for model M1.5b0.270 at different resolutions for the finest
  grid.  The straight line represents the result of the fit (as a
  function of $\beta$) for the whole sequence of models at fixed value
  of the conserved baryonic mass $M_0=1.5 M_\odot$ and at resolution
  $dx=0.50$. The open triangles represent the values of two
  different models among them.
\label{fig:tau_resolution}}
\end{figure}

While the overall dynamics is very
similar for all resolutions, there is a consistent shift of the value of the
$\beta$ parameter at the beginning of the development of the instability
$\beta_i$ and of the value of the fitted growth time $\tau_2$ with increasing
resolution.
More precisely, the analysis of the initial stage of the evolution shows
(see Fig.~\ref{fig:beta_evolution_res}) that there is indeed a drift (decrease) of the
value of $\beta$ and, consequently, the fit of $\tau_2$ is sensible to the value used
for a given resolution. This can, at least partially, be explained to be due to the fact that
by the time the amplitude of the mode reaches the fit-window, the evolution
does not correspond any longer to the original model but it is closer to a new 
equilibrium model (through an adiabatic drift), characterized by a different value
of the rotational parameter $\beta$.
The overall conclusion is that one has to be careful when associating
the fitted value for the growth time of the instability to the initial value of the
rotational parameter $\beta$. In fact, as it can be seen in 
Fig.~\ref{fig:beta_evolution_res}, there is a small shift (of as much as 2\% for
the lowest resolution)
of the value of $\beta$ from the start of the simulation up to the time
at which the instability is detected ($\beta_i=\beta(t_i)$). 
Using this fact, we now report in Fig.~\ref{fig:tau_resolution} the growth-time 
and the $\beta$ at the beginning of the instability for each resolution and 
the critical fit of Eq.~(\ref{eq:res_fits}) for $M_0=1.5 M_\odot$.
That shows we have convergence in the determination of the parameter of the
classical bar-mode instability above the threshold, if we use the 
value $\beta_i$ to perform the extrapolations.

\begin{figure}
\vspace{-1mm}
\begin{centering}
  \includegraphics[width=0.45\textwidth]{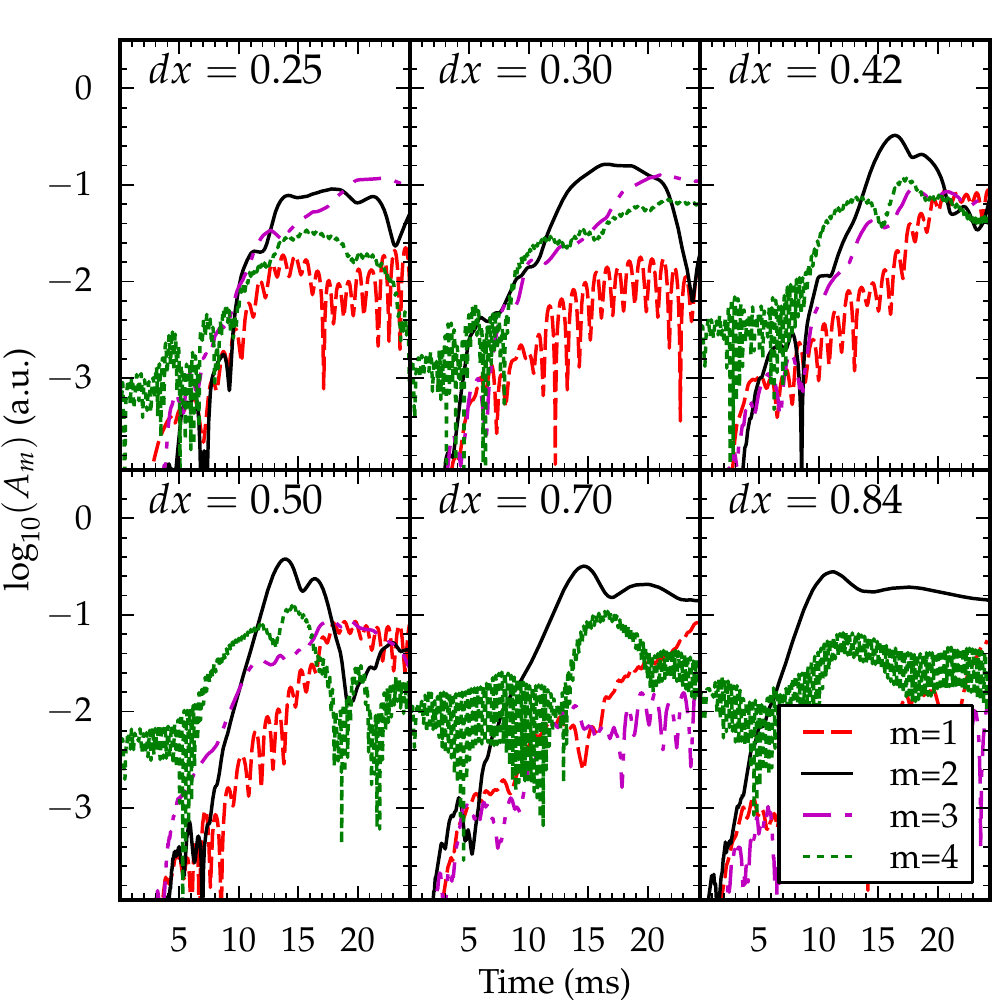}
\end{centering}
\vspace{-4mm}
\caption{Resolution's effect on the mode dynamics for model M2.5b0.200. In the case
of models below the threshold the interplay of modes with different shapes
does not show a clear convergent behavior on the dynamics. 
\label{fig:M25b200res}}
\end{figure}

While these results show that the used resolution is enough to
explore the dynamical bar-mode instability,
we cannot draw such conclusion on the 
parameter of the observed shear-instability at lower $\beta$. 
Despite that we did observe that such instabilities are present, the resolution used here
is not sufficient to draw conclusions about their numerical 
values. We note that we found candidates of $m=2$ shear
instability when examining several models in different resolutions,
but the instability dynamics cannot be clearly singled out with
respect to other instabilities that seem to be present. One example
is given in Fig.~\ref{fig:M25b200res}, which shows that the interplay on the growth of
various modes shows some dependency on the resolution. Such dependency was not observed in 
the cases where the classical dynamical bar-mode instability is present 
and it is by far the fastest growing mode.

This shows that for these values of $\beta$ the used resolution on the finer grid
of $dx=0.50$ is not enough to determine the dynamics
of the shear instability. This initial analysis show that a resolution of, at
least, $dx=0.25$ is needed. 
Moreover, in this case, the use of the numerical discretization error 
to triggering the fastest growing mode does not seem to be the best 
strategy to study shear instabilities 
when a competition between different $m$ modes (like $m=2$ and $m=3$) is present.
We will leave a detailed study of the low-$beta$ instabilities to a future 
work.

\section{Conclusions}
\label{sec:conclusions}

We have presented a study of the dynamical bar-mode instability in
differentially rotating NSs in full General Relativity for a wide and
systematic range of values of the rotational parameter $\beta$ and the
conserved baryonic mass $M_0$, using a polytropic/ideal-fluid EOS
characterized by a value of the adiabatic index $\Gamma=2.75$, which
allows us to resemble the properties of a realistic EOS.
In particular, we have evolved a large number of NS models belonging
to five different sequences with a constant rest-mass
ranging from $0.5$ to $2.5 \, M_\odot$, with a fixed degree of
differential rotation ($\hat{A} = 1$) and with many different values
of $\beta$ in the range $[0.140,0.272]$.

For all the models with a sufficiently high initial value of $\beta$
we observe the expected exponential growth of the $m=2$ mode which is
characteristic of the development of the dynamical bar-mode
instability.  We compute the growth time $\tau_2$ for each of these
bar-mode unstable models by performing a nonlinear least-square fit
using a trial function for the quadrupole moment of the matter
distribution.  The growth time clearly depends on both the rest-mass
and the rotation and in particular we find that the relation between
the instability parameter $\beta$ and the inverse square of $\tau_2$,
for each sequence of constant rest-mass, is linear.  

This allows us to extrapolate the threshold value $\beta_c$ for each sequence
corresponding to the growth time going to infinity, using the same
procedure already employed in \cite{Manca:2007ca}. Once the five values of $\beta_c$
have been computed, we are able to extrapolate the critical value of
the instability parameter for the Newtonian limit, which is found to
be $\beta_c^N |_{\Gamma = 2.75} = 0.2527$.  This value can be directly
compared with the one found in \cite{Manca:2007ca} for the ``standard'' $\Gamma = 2$
case, which is $\beta_c^N |_{\Gamma = 2}=0.266$.

Our results suggest that, even if one can now consider just two values for
the adiabatic index, namely the values $\Gamma = 2.75$, considered in
the present work, and $\Gamma = 2$, considered in
\cite{Baiotti:2006wn,Manca:2007ca}, the use of a stiffer, more
realistic EOS should be expected to have the effect of reducing the
threshold for the onset of the dynamical bar-mode instability.
Unfortunately, the actual reduction of the threshold $\beta_c$ is just
of the order of $5\%$, and indeed this reduction does not lead to a
significantly higher probability for it to occur in real astrophysical scenarios.

We also evolved many models belonging to the same five sequences
but having lower values of the instability parameter $\beta$. 
We find that many of them show the growth of one or more modes 
even though their initial value of $\beta$ is below the threshold
for the onset of the dynamical bar-mode instability. The modes 
that show a growth are mainly $m=2$ and the $m=3$.
We compute the frequencies of these growing modes and compare them
with the corotation band for their progenitor models, finding
that all those frequencies are within this band.
We can conclude
that such instabilities have to be defined as \textit{shear instabilities},
as the ones that were already observed in \cite{Corvino:2010}.

Unfortunately, we are not able to measure their growth time, since 
their dynamics change significantly by changing the resolution of
the simulations. In fact, while at a coarse resolution we usually 
observe only one mode growing exponentially, when improving the resolution
other modes develop as well and the interplay between these prevent a clear
exponential growth of only one mode which could dominate the evolution.

In order to 
make a quantitative assessment about this phenomenon, either much
higher resolution has to be used to see if one of the modes is able to
dominate, or seed perturbations have to be introduced
with the aim of selecting only a particular mode at a time.
We leave this treatment to futures studies.

\acknowledgments

We do have to especially thank N.~Stergioulas for providing us the RNS code that we
used to generate the initial stellar configurations. We would also like to thank 
R.~Alfieri, S.~Bernuzzi, N.~Bucciantini, A.~Nagar, L.~Del~Zanna, 
for useful discussions and insights in the development of the present
work. Portions of this research were conducted with high performance computing (HPC)
resources provided by the European Union PRACE program (6$^{th}$ call, project ``3DMagRoI''),
by the Louisiana State University (allocations hpc\_cactus, hpc\_numrel and hpc\_hyrel), 
by the Louisiana Optical Network Initiative (allocations loni\_cactus and loni\_numrel);
by the National Science Foundation through XSEDE resources (allocations TG-ASC120003, 
TG-PHY100033 and TG-MCA02N014), by the INFN ``Theophys'' cluster and through the 
allocation of CPU time on the BlueGene/Q-Fermi at CINECA for the specific 
initiative INFN-OG51 under the agreement between INFN and CINECA. The work of A.~F. has 
been supported by MIUR (Italy) through the INFN-SUMA project. F.~L. is directly
supported by, and this project heavily used infrastructure developed using support
from the National Science Foundation in the USA (1212401 / 1212426 / 1212433 / 1212460).

\end{document}